\documentclass[aps,prd,superscriptaddress,amsmath,twocolumn,10pt,nofootinbib]{revtex4}

\usepackage{color,hangcaption,hhline,psfrag,rotating,amssymb}
\usepackage[hang,nooneline]{subfigure}
\usepackage{dcolumn}
\usepackage{verbatim}
\usepackage{latexsym}

\newcommand{\Tr}{\mathrm{Tr}}
\newcommand{\On}{\mathcal{O}^{(n)}}
\newcommand{\psib}{\bar{\psi}}
\newcommand{\eq}[1]{Eq.\:(\ref{#1})}
\newcommand{\cond}{M\psib\psi}
\newcommand{\mcond}{m\psib\psi}

\newcommand{\vacexp}[1]{\langle 0 | #1 | 0\rangle}
\newcommand{\qexp}[1]{\langle q | #1 | q^\prime\rangle}
\newcommand{\Mb}{\bar{M}}
\newcommand{\mb}{\bar{m}}
\newcommand{\alphab}{\bar{\alpha}}
\newcommand{\msb}{{\overline{\mathrm{MS}}}}

\begin{document}
\title{Direct determination of the strange and light quark condensates from full lattice QCD}

\author{C. McNeile}
\email[]{mcneile@uni-wuppertal.de}
\affiliation{Bergische Universit\"{a}t Wuppertal, Gaussstr.\,20, D-42119 Wuppertal, Germany}
\author{A. Bazavov}
\affiliation{Physics Department, Brookhaven National Laboratory, Upton NY 11973, USA}
\author{C.~T.~H.~Davies}
\email[]{c.davies@physics.gla.ac.uk}
\affiliation{SUPA, School of Physics and Astronomy, University of Glasgow, Glasgow, G12 8QQ, UK}
\author{R.~J.~Dowdall}
\affiliation{SUPA, School of Physics and Astronomy, University of Glasgow, Glasgow, G12 8QQ, UK}
\author{K. Hornbostel}
\affiliation{Southern Methodist University, Dallas, Texas 75275, USA}
\author{G. P. Lepage}
\affiliation{Laboratory of Elementary-Particle Physics, Cornell University, Ithaca, New York 14853, USA}
\author{H. D. Trottier}
\affiliation{Physics Department, Simon Fraser University, 8888 University Drive, Burnaby, BC, V5A 1S6, Canada}
\affiliation{TRIUMF, 4004 Westbrook Mall, Vancouver, BC, V6T 2A3, Canada}


\date{\today}

\begin{abstract}
We determine the strange quark condensate from 
lattice QCD for the first time and compare its value 
to that of the light quark and chiral condensates. 
The results come from a direct calculation of the expectation value 
of the trace of the quark propagator followed by subtraction of the 
appropriate perturbative contribution, derived here, to convert 
the non-normal-ordered $m\overline{\psi}\psi$ 
to the $\overline{MS}$ scheme at a fixed 
scale. This is then a well-defined physical `nonperturbative' condensate 
that can be used in the Operator Product Expansion of current-current 
correlators. 
The perturbative subtraction is calculated 
through $\mathcal{O}(\alpha_s)$ and
estimates of higher order terms are included through fitting 
results at multiple lattice spacing values. 
The gluon field configurations used are `second generation' 
ensembles from the MILC collaboration
that include $2+1+1$ flavors of sea quarks 
implemented 
with the Highly Improved Staggered Quark action and including $u/d$ 
sea quarks down to physical masses.  
Our results are : 
$\langle \overline{s}{s} \rangle^{\overline{MS}}(2\, \mathrm{GeV})= -(290(15)\, \mathrm{MeV})^3$, 
$\langle \overline{l}{l} \rangle^{\overline{MS}}(2\, \mathrm{GeV})= -(283(2)\, \mathrm{MeV})^3$, 
where $l$ is a light quark with mass equal to the average of 
the $u$ and $d$ quarks. The strange to light quark condensate ratio is 1.08(16).  
The light quark condensate is significantly larger than 
the chiral condensate in line with expectations from chiral analyses.
We discuss the implications of these results for other calculations. 
\end{abstract}


\maketitle

%
\section{Introduction}
\label{sec:intro}
A critical feature of the nonperturbative dynamics of QCD 
at zero temperature
is the condensation of quark-antiquark pairs in the 
vacuum, spontaneously breaking the chiral symmetry of the action. 
The value of the chiral condensate (the 
quark condensate at zero quark mass) is then an important  
parameter for low energy QCD~\cite{Reinders:1984sr}. The well-known Gell-Mann, Oakes, 
Renner (GMOR) relation~\cite{GellMann:1968rz}:
\begin{equation}
\frac{f_{\pi}^2M_{\pi}^2}{4} = - \frac{m_u+m_d}{2}\frac{\langle 0 | \overline{u}u + \overline{d}d | 0 \rangle}{2}
\label{eq:gmor} 
\end{equation}
connects the $u/d$ quark masses times 
condensate to the square of the mass times decay constant for the  
Goldstone boson of the spontaneously broken symmetry. 
Eq.(~\ref{eq:gmor}) has normalisation such that $f_{\pi} = 130$ MeV. 
The GMOR 
relation holds in the limit of $m_u, m_d \rightarrow 0$. 
A value for this chiral condensate can be derived from the chiral 
extrapolation of lattice 
QCD results for light meson masses and decay constants. 
See, for example, the recent result of 
$-(272(2)\,\mathrm{MeV})^3$ for the chiral 
condensate 
in the $\overline{MS}$ scheme at 2 GeV
using SU(2) chiral perturbation theory~\cite{Borsanyi:2012zv}. 

The determination of the quark condensate for non-zero
quark masses is more problematic because, depending 
on the method used, there are various sources of unphysical 
quark mass dependence and a careful definition of the 
condensate is required. This definition must be phrased 
in terms of the Operator Product Expansion (OPE) since this 
is the context in which the condensate appears~\cite{Shifman:1978by, Reinders:1984sr, Shifman:1998rb}. The OPE 
allows us to separate short and long-distance contributions 
in, for example, a short-distance current-current 
correlator. The expansion is in terms of a set 
of matrix elements of local operators 
multiplied by coefficient functions. The aim is for all the long-distance 
contributions (with scale $< \mu$) to be contained in the matrix elements and 
the short distance contributions (with scale $> \mu$) 
in the coefficient functions. A key matrix element, since it 
corresponds to a relatively low-dimensional ($d=3$) operator, is that of
the quark condensate.  The clean separation of scales in 
the OPE only works if the local operators are {\it not} normal 
ordered~\cite{Chetyrkin:1985kn, Jamin:1992se}. Then the coefficient 
functions are analytic in the quark masses and therefore free 
of infrared sensitivity. This means, however, that the 
quark mass dependent mixing of $m\overline{\psi}\psi$ with the unit operator 
must be taken into account and that the vacuum matrix 
element of $m\overline{\psi}\psi$ is not cut-off independent.  
The quantity that appears in the OPE is the vacuum 
matrix element in, for example, the $\overline{MS}$ scheme 
at the scale $\mu$.  We can derive this matrix element 
from lattice QCD and we give results here for $\mu$ = 2 GeV. 
The results can easily be run to other scales, as appropriate. 

The value of the condensate for quarks of non-zero 
mass up to that of the strange quark is needed in a number of calculations
involving light quark correlators. In lattice QCD 
it is frequently easier and statistically more precise 
to use strange quarks than very light 
quarks in contexts where the quark mass is not expected 
to be important. Then the 
strange quark condensate appears in the calculation, however. 
Examples include the matching to continuum QCD 
perturbation theory of lattice QCD calculations of moments of heavy-light 
meson correlators~\cite{Koponen:2010jy} 
and of light meson correlators at large 
space-like $q^2$~\cite{Shintani:2010ph}. 
Such calculations are used to extract quark masses and the strong 
coupling constant, $\alpha_s$. 
A continuum example where the strange 
quark condensate is needed is in the determination of 
the strange quark mass, $m_s$, from hadronic $\tau$ decays~\cite{Gamiz:2002nu}. 

Current estimates of the value of 
the strange quark condensate vary by almost 
a factor of two~\cite{Albuquerque:2009pr, Maltman:2008na}.
It is not even clear 
whether the strange condensate is larger or smaller than the light 
quark condensate. For very large quark masses, $m_q > \Lambda_{QCD}$, say, 
so that the quark mass dominates the propagator,
it seems clear that the 
condensate should fall to zero, but this does not help in 
determining the slope of the condensate with $m_q$ for small 
quark masses.  

Here we address the determination of the strange 
(or other non-zero mass) 
quark condensate by direct calculation in full lattice QCD. By direct 
we mean that we determine the vacuum expectation value of the 
strange quark propagator as well as the light quark propagator 
on a range of gluon field configurations at different values of 
the lattice spacing and sea quark masses. 
To isolate the low-energy nonperturbative 
value of the condensate from these results requires the subtraction 
of a perturbative contribution. 
The perturbative 
contribution in lattice QCD has two pieces. One diverges as 
$a \rightarrow 0$ and dominates the vacuum expectation value of 
the strange quark propagator, particularly on our finer lattices. 
The second piece contains infrared sensitive logarithms of the 
quark mass which cancel against similar terms in continuum 
perturbation theory allowing an infrared safe definition of 
the condensate for use in the OPE, as discussed above. 

The error in the final result then depends on how well this 
subtraction can be done. Here we use an explicit calculation 
of the perturbative pieces through $\mathcal{O}(\alpha_s)$
and fit for unknown higher order terms. The known quark mass 
and $a$ dependence of these unknown terms helps in constraining 
them along with the very small statistical errors in our lattice 
results. We also use a particularly good discretisation of 
the Dirac action known as the Highly Improved Staggered Quark (HISQ) 
formalism~\cite{Follana:2006rc} 
on `second generation' gluon field configurations so that 
discretisation errors in the physical nonperturbative results are small. 

The paper is laid out as follows. In Section~\ref{sec:theory} we 
describe the theoretical background to direct calculations 
of the quark condensate in lattice QCD. Section~\ref{sec:lattice}
gives our lattice QCD results on gluon configurations with 
2+1+1 flavors of sea quarks, describing the calculation of 
the perturbative contribution that is subtracted and then 
the procedure for fitting the remaining nonperturbative condensates
as a function of quark mass and lattice spacing. 
We also give results from configurations including 2+1 flavors 
of sea quarks over a wider range of lattice spacing values but 
studying only the strange quark condensate in Appendix~\ref{app:lattice2}. 
In Section~\ref{sec:discussion} we compare to previous 
values and discuss the implications 
of our results for both zero and finite temperature QCD calculations. 
Section~\ref{sec:conclusions} gives our conclusions. 

%
\section{Theoretical Background}
\label{sec:theory}
The direct determination of the chiral condensate in lattice QCD
requires the calculation of the 
expectation value over an ensemble of 
gluon fields, $U$, of $\mathrm{Tr}M^{-1}$ where $M$ 
is the lattice discretisation of the Dirac matrix. 
The quark action for a given quark flavor, 
\begin{equation}
S_f = \overline{\psi} M_f \psi
\end{equation} 
and 
\begin{equation}
\langle \overline{\psi} \psi \rangle = \langle 0 | \overline{\psi}_f \psi_f | 0 \rangle = -\frac{1}{V}\langle \mathrm{Tr} M_f(U)^{-1}\rangle_U ,
\label{eq:pptrace}
\end{equation}
where the trace is over spin, color and space-time point 
and the gluon fields in the ensemble used for 
the average include the effect 
of sea quarks (of all flavors, not just $f$) in their probability distribution. 
$V$ is the lattice volume, $L^3\times T$. 
For a naive discretisation of the Dirac action $M$ takes the 
form: 
\begin{equation}
M = \gamma_{\mu}\Delta_{\mu} + m
\label{eq:Mdef}
\end{equation}
where $\Delta_{\mu}$ is a covariant finite 
difference on the lattice: 
\begin{equation}
\Delta_{\mu} \psi_x = \frac{1}{2a} \left( U_{\mu}(x)\psi(x+\hat{\mu}) - U_{\mu}^{\dagger}(x-\hat{\mu})\psi(x-\hat{\mu}) \right)
\label{eq:deltadef}
\end{equation}
and $m$ is the quark mass for that flavor.  Because of 
fermion doubling this formalism describes 
16 `tastes' of quarks in 4-dimensions rather 
than just 1 and we must divide the right-hand 
side of Eq.~(\ref{eq:pptrace}) by the number 
of tastes, $N_t=16$.
The staggered formalism is derived from this naive formalism 
by a rotation which allows the spin degree of freedom to be dropped. 
In that case the quark 
field becomes a 1-component spinor, which is numerically very efficient, and $N_t=4$. 

For $m=0$ the eigenvalues of $M$ for either naive or staggered 
quarks, are purely imaginary and come 
in $\pm$ pairs. Therefore, in the absence of exact zero modes, 
\begin{eqnarray}
- \langle \overline{\psi}\psi \rangle &=& \frac{1}{N_t}\sum_{\lambda} \left(\frac{1}{m+i\lambda} + \frac{1}{m-i\lambda} \right) \nonumber \\
&=& \frac{1}{N_t} \sum_{\lambda} \frac{2m}{m^2 + \lambda^2}. 
\label{eq:ppeigs}
\end{eqnarray}
A calculation at 
$m=0$ on a finite volume lattice would then give an answer for the 
quark condensate of zero. This does not 
mean that chiral symmetry is unbroken, however. 
The problem arises because the broad distribution of non-zero eigenvalues 
(ignoring topological near-zero modes) drops to zero near 
the origin in a way that depends strongly on the volume. 
Once the quark mass is below the minimum of the non-zero 
eigenvalues the result for $\langle \overline{\psi}\psi \rangle$ 
will be distorted. 
A more careful consideration 
of limits must be made. If $V$ is taken to infinity before $m$ is 
taken to zero then the sum over eigenvalues can be replaced by an 
integral and the Banks-Casher relation~\cite{Banks:1979yr} is obtained. This connects 
the zero-mass condensate to the spectral density at the origin: 
\begin{equation}
\Sigma = - \langle \overline{\psi}\psi(m=0) \rangle = \frac{\pi \rho(0)}{N_t}. 
\label{eq:banks-casher}
\end{equation}
Thus $\Sigma$ can be obtained from studies of the spectral density 
and, more recently, has also been obtained from matching the 
distribution of low eigenmodes to random matrix theory 
in the $\epsilon$ regime~\cite{Leutwyler:1992yt, Shuryak:1992pi, Damgaard:2001ep}. 
Here we are more concerned with extracting a condensate at non-zero 
quark mass, for example at the strange quark mass, and so the 
issue above is not relevant. We will work on large volume lattices 
(over ten times larger than the study in~\cite{Follana:2005km} that looked 
at staggered eigenvalues in the $\epsilon$ regime) 
at quark mass values that are well within the distribution 
of non-zero eigenvalues. Our results for $\langle \mathrm{Tr} M^{-1} \rangle$
then include both the effects of a non-zero value for $\rho(0)$ and 
a non-zero quark mass. 

A second issue arises, however, in the extraction of a physical 
nonperturbative value for the condensate at non-zero quark mass. 
A perturbative contribution appears from mixing between  
the scalar $\overline{\psi}\psi$ operator and the identity since 
the identity operator has a vacuum expectation value in 
the trivial perturbative vacuum. 
This perturbative contribution vanishes at zero quark mass since 
chiral symmetry is not broken in perturbation theory (for the 
same reason as that given on the lattice above in Eq.~(\ref{eq:ppeigs})). 
At non-zero quark mass it contains odd powers of $m$ 
starting with a quadratically ultra-violet divergent term linear in $m$. 
This can be illustrated with a tree-level calculation in 
the continuum to give: 
\begin{eqnarray}
-\langle \overline{\psi}\psi \rangle &=& \int_0^{\Lambda} \frac{d^4k}{(2\pi)^4}\frac{12m}{k^2+m^2} \\
&=& \frac{3}{4\pi^2} \left( m\Lambda^2 + m^3 \log \frac{m^2}{\Lambda^2+m^2}\right) . \nonumber
\label{eq:contdiv}
\end{eqnarray}
The quadratic ultraviolet divergence depends on the 
scheme used but the 
$m^3 \log(m/\Lambda)$ term is universal since it arises from 
the infrared part of the integral. The coefficient above agrees 
with that obtained for the $\overline{MS}$ scheme in~\cite{Braaten:1991qm} 
and on the lattice for highly improved staggered quarks 
to be described in section~\ref{subsec:pt}. 
We stress that a perturbative contribution of this kind is present for 
all lattice regularisations of QCD, whatever their chiral symmetry 
properties, and so must be calculated and subtracted to give a 
physical result. The quadratic divergence present for Ginsparg-Wilson 
fermions is demonstrated in the quenched approximation in, for 
example,~\cite{Hernandez:1999cu} and the additional divergences for 
the Wilson formalism with broken chiral symmetry in~\cite{David:1984ys}.  

This subtraction is somewhat analogous to subtracting perturbative 
contributions to the mean plaquette to obtain the nonperturbative 
gluon condensate. That, however, is extremely difficult to do 
because the nonperturbative condensate contribution 
to the plaquette is so small. This contribution is given at leading order by:
\begin{equation}
\delta P_{cond} = -\frac{\pi^2}{36} a^4 \langle \alpha_s G^2/\pi \rangle .
\label{eq:plaqcond}
\end{equation}
If we take the value of the gluon condensate as $\mathcal{O}(\Lambda_{QCD}^4)$ 
then $\langle \alpha_s G^2 / \pi \rangle \approx 0.005 \,\mathrm{GeV}^4$. 
On very coarse lattices for which 
$a \approx 1 \,\mathrm{GeV}^{-1}$, this contributes less than 1\% to the 
value of the plaquette. On finer lattices the nonperturbative 
condensate contribution is even smaller because it falls as $a^4$ 
while the perturbative contribution falls only as $\alpha_s(d/a)$ for 
some scale $d$. This means that the plaquette is in fact a very 
good variable to use for the determination of $\alpha_s$ from 
lattice QCD calculations but not for the determination of the 
gluon condensate~\cite{Davies:2008sw}. For larger Wilson loops the gluon condensate 
contribution is larger, being proportional to the square of the 
area of the loop, but the coefficients in the perturbative series 
also become larger.  

The determination of the nonperturbative quark condensate from 
$\langle \mathrm{Tr} M^{-1} \rangle$ is in much better shape 
than this for several reasons. The main one is that the 
nonperturbative condensate contribution to 
$\langle \mathrm{Tr} M^{-1} \rangle$ 
in lattice units is only a factor of $a^2$ smaller than 
the leading perturbative contribution rather than $a^4$. 
In addition the perturbative contribution is suppressed 
by the quark mass, which is small for the $u/d$ and $s$ 
quarks we will consider here. The perturbative contribution 
is a well-defined function of the quark mass at every 
order in perturbation theory and so results at several 
values of the quark mass, and the lattice spacing, can 
be used to constrain unknown higher orders, beyond 
the $\mathcal{O}(\alpha_s)$ that we have explicitly 
calculated, and we will 
make use of that here. 

\section{Lattice QCD calculation on $n_f=2+1+1$ gluon configurations}
\label{sec:lattice}

The gluon field configurations used here are listed 
in Table~\ref{tab:params}. They were generated by 
the MILC collaboration~\cite{Bazavov:2010ru} using a 
tadpole-improved L\"{u}scher-Weisz gauge action with coefficients 
corrected perturbatively through $\mathcal{O}(\alpha_s)$ 
including pieces proportional to $n_f$, the number 
of quark flavors in the sea~\cite{Hart:2008sq}.
The gauge action is then improved completely 
through $\mathcal{O}(\alpha_sa^2)$. 
Sea quarks are included with the highly improved 
staggered quark (HISQ) action~\cite{Follana:2006rc} 
which has been designed to have very small 
discretisation errors. 
Discretisation errors are formally removed through $\mathcal{O}(a^2)$ 
but higher order errors, particularly staggered taste-changing 
errors, are seen to be smaller with HISQ than with the earlier 
asqtad staggered quark action~\cite{Follana:2006rc, Bazavov:2010ru}. 
The HISQ action used here has two smearing steps for the gluon 
field appearing in the quark action with a U(3) projection 
of the smeared links between the two smearing steps.
The configurations include a sea
charm quark in addition to up, down and strange. 
These configurations are then said to have 2+1+1 flavors in 
the sea, since the $u$ and $d$ quarks are taken to have the 
same mass (denoted $m_l$ here). This is heavier than the average $u/d$ mass 
in the real world on most of the configuration sets but there 
are three for which the $u/d$ mass has its physical 
value (3, 6 and 8). The 
$s$ and $c$ masses are tuned as closely as possible to 
their correct values on each set. The tuning of the 
sea $s$ quark mass is accurately done -- typically to better 
than 5\%  -- so 
the $u/d$ quark mass can 
be accurately calibrated in terms of the $s$ quark 
mass for chiral extrapolations. 
The values of the lattice spacing for most ensembles were 
determined in~\cite{Dowdall:2011wh} using the decay constant of 
the $\eta_s$ meson. The values vary from 0.15 fm to 0.09 fm 
as we go from the very coarse to the fine lattices. 
The spatial volumes are large, from $(2.5 \,\mathrm{fm})^3$ 
when $m_l/m_s \approx 0.2$ to $(3.7 \,\mathrm{fm})^3$ when 
$m_l/m_s \approx 0.1$.  

\begin{table}
\caption{
Details of the MILC gluon field ensembles used in this paper. 
$\beta=10/g^2$ is the $SU(3)$ gauge coupling and $L/a$ and $T/a$ 
are the number of lattice spacings in the space and 
time directions for each lattice. 
$am_{l,sea},am_{s,sea}$ and $am_{c,sea}$ are the light (up and down taken to 
have the same mass), strange and charm sea quark masses in lattice units.
$a$ is the lattice spacing in fm determined from the decay constant 
of the $\eta_s$ meson in~\cite{Dowdall:2011wh} with values for 3, 6 and 8 added 
here.
The ensembles 1, 2 and 3 will be referred to in the text as ``very coarse'', 
4, 5 and 6 as ``coarse'' and 7 and 8 as ``fine.'' 
}
\label{tab:params}
\begin{ruledtabular}
\begin{tabular}{llllllll}
Set & $\beta$ & $a$/fm & $am_{l,sea}$ & $am_{s,sea}$ & $am_{c,sea}$ & $L/a \times T/a$ \\
\hline
1 & 5.80 & 0.1546(11) & 0.013   & 0.065  & 0.838 & 16$\times$48 \\
2 & 5.80 & 0.1526(8) & 0.0064  & 0.064  & 0.828 & 24$\times$48 \\
3 & 5.80 & 0.1511(8) & 0.00235  & 0.0647  & 0.831 & 32$\times$48 \\
\hline
4 & 6.00 & 0.1234(8)  & 0.0102  & 0.0509 & 0.635 & 24$\times$64 \\
5 & 6.00 & 0.1218(6) & 0.00507 & 0.0507 & 0.628 & 32$\times$64 \\
6 & 6.00 & 0.1206(6) & 0.00184 & 0.0507 & 0.628 & 48$\times$64 \\
\hline
7 & 6.30 & 0.0899(7)  & 0.0074  & 0.0370  & 0.440 & 32$\times$96 \\
8 & 6.30 & 0.0875(7) & 0.0012 & 0.0363 & 0.432 & 64$\times$96 \\
\end{tabular}
\end{ruledtabular}
\end{table}

\begin{figure}
\includegraphics[width=0.9\hsize]{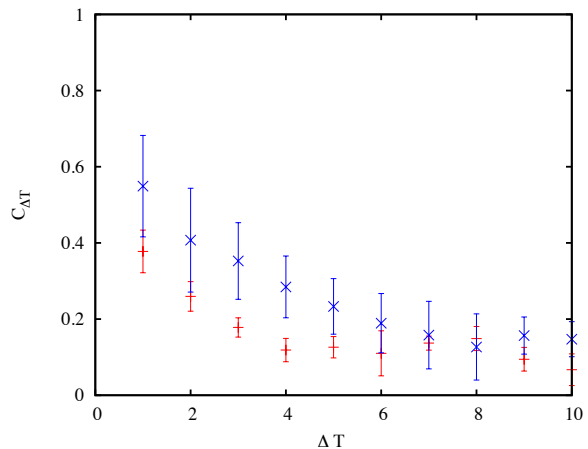}
\caption{
Autocorrelation function, $C_{\Delta T}$ of Eq.~\ref{eq:autocorr}, 
for the strange and light quark condensates on coarse set 6 
with physical mass light sea quarks. 
The $x$-axis, $\Delta T$, is the separation in time units 
between configurations. The strange condensate results are 
given as blue crosses and the light condensate with 
red pluses. Errors in $C_{\Delta T}$ are estimated 
by dividing the configuration time series into five 
consecutive sets. 
}
\label{fig:autocorr}
\end{figure}

\begin{table*}
\caption{
Raw (unsubtracted) values for the 
light and strange quark condensates in lattice 
units calculated for valence masses given in columns 2 and 3. 
The results use the correlators 
calculated in~\cite{Dowdall:2011wh} (via 
Eq.~(\ref{eq:ppbardef})), but we also give results for 
additional strange quark masses on sets 1 and 2 and new results 
on sets 3, 6 and 8. We have 16,000 correlators per 
ensemble, except for sets 6 and 8 where we use approximately
10000. 
}
\label{tab:valence}
\begin{ruledtabular}
\begin{tabular}{lllllllll}
Set & $am_{l,val}$ & $aM_{\pi}$ & $af_{\pi}$ & -$a^3 \langle \overline{\psi} \psi_l \rangle_{0}$ & $am_{s,val}$ & $aM_{\eta_s}$ & $af_{\eta_s}$ & -$a^3 \langle \overline{\psi} \psi_s \rangle_{0}$ \\
\hline
1 & 0.013 & 0.23637(15) & 0.11183(9) & 0.018607(29) & 0.0688 & 0.53361(14) & 0.14199(6) & 0.045758(19) \\
  & & & & & 0.0641 & 0.51491(14) & 0.13996(6) & 0.043616(19) \\
2 & 0.0064 & 0.16615(7) & 0.10511(5) & 0.014524(18) & 0.0679 & 0.52797(8) & 0.14026(3) & 0.045009(12) \\
  & & & & & 0.0636 & 0.51078(8) & 0.13839(3) & 0.043038(12) \\
3 & 0.00235 & 0.10172(5) & 0.09934(5) & 0.011762(11) & 0.0628 & 0.50657(5) & 0.13720(3) & 0.042483(6) \\
\hline
4 & 0.01044 & 0.19153(9) & 0.09075(5) & 0.011629(13) & 0.0522 & 0.42351(9) & 0.11312(4) & 0.031756(10) \\
5 & 0.00507 & 0.13413(5) & 0.08451(4) & 0.008511(9) & 0.0505 & 0.41476(6) & 0.11119(2) & 0.030768(6) \\
6 & 0.00184 & 0.08154(2) & 0.07988(2) & 0.006534(6) & 0.0507 & 0.41481(2) & 0.11062(2) & 0.030768(4) \\
\hline
7 & 0.0074 & 0.14070(9) & 0.06621(5) & 0.006153(8) & 0.0364 & 0.30884(11) & 0.08238(4) & 0.019822(4) \\
8 & 0.0012 & 0.05718(2) & 0.05781(3) & 0.002803(4) & 0.0360 & 0.30483(4) & 0.08055(2) & 0.019504(2) \\
\end{tabular}
\end{ruledtabular}
\end{table*}

On each of these ensembles we determine $\langle \mathrm{Tr} M^{-1} \rangle$
for HISQ valence quarks for various quark masses. 
To do this we use an identity that relates the quark propagator for 
staggered quarks to a product of quark propagators: 
\begin{eqnarray}
\frac{1}{am_q} \mathrm{Tr} M^{-1}_{00} &=& \sum_n \mathrm{Tr}\left[ M^{-1}_{0n}M^{-1}_{n0}\right] (-1)^n \nonumber \\
&=& \sum_n \mathrm{Tr} |M^{-1}_{0n}|^2 
\label{eq:condid}
\end{eqnarray} 
Here 0 and $n$ are arbitrary lattice sites and $am_q$ is the quark mass 
in lattice units used for the quark propagator. 
The righthand side of Eq.~(\ref{eq:condid}) is simply 
the correlator between 0 and $n$ for the Goldstone pseudoscalar meson made of 
a quark and antiquark of mass $am_q$. Summing over $n$ projects 
on to zero spatial momentum and sums over timeslices. Thus, dividing 
both sides by 4, the number of tastes for staggered quarks, we obtain:
\begin{equation}
-a^3 \langle \overline{\psi} \psi \rangle_0 = (am_q) \sum_t C_{\pi}(t).
\label{eq:ppbardef}
\end{equation}
The raw condensate value on the left-hand side of this expression 
is normalised to the single flavor case and the pion correlator on 
the right-hand side is the usual zero-momentum Goldstone meson correlator.   
This allows us to determine $\langle \overline{\psi} \psi \rangle_0$ by 
summing over 
the Goldstone pseudoscalar correlators calculated in~\cite{Dowdall:2011wh}. 
Eq.~(\ref{eq:condid}) is derived in~\cite{Kilcup:1986dg} for a single propagator 
origin, 0, but the derivation can trivially be extended to hold for 
the random wall source that we use for our correlators in~\cite{Dowdall:2011wh}. 
The identity holds configuration by configuration for lattice QCD 
quark formalisms with sufficient chiral symmetry and in the continuum 
for a specific gauge field background. We give an explicit proof of 
this in Appendix~\ref{appendix:appa}. 
Since our Goldstone pseudoscalar correlators are sums of positive 
numbers they are particularly precise and this precision then 
carries over to our condensate results. 
For the light condensate we use the pion correlators made of light quarks 
and for the strange condensate we use the $\eta_s$ correlators made 
of strange quarks. 
We stress that what we calculate using Eq.~(\ref{eq:condid}) 
is the vacuum 
expectation value of the condensate (and not a specific `in-meson' value) 
despite the fact that we determine it for convenience from a meson correlator. 

Table~\ref{tab:valence} gives the valence quark masses used in our calculation
and the raw results for the condensate obtained from Eq.~(\ref{eq:condid}). 
The correlator calculations used 16 `random wall' time sources 
on approximately one thousand 
configurations in each ensemble (somewhat fewer on sets 6 and 8)
and so the results have very small statistical 
errors. The valence light quark masses are equal to those in sea (except 
for a very small change on set 3) but we have shifted the valence strange 
quark masses slightly to be closer to the physical strange quark 
mass, following~\cite{Dowdall:2011wh}. On sets 1 and 2 we give results for two 
different values of the strange quark mass, to help in constraining the 
valence mass dependence of the condensate.   

Errors on the condensate values are determined after binning 
over adjacent sets of at least five configurations, following 
analysis of the autocorrelation function. An example plot is 
shown, for coarse set 6, in Fig.~\ref{fig:autocorr}. 
The autocorrelation function is defined as: 
\begin{equation}
C_{\Delta T} = \frac{\langle x_i x_{i+\Delta T} \rangle - \langle x_i \rangle \langle x_{i+\Delta T} \rangle}{\langle x_i^2 \rangle - \langle x_i \rangle^2 } .
\label{eq:autocorr}
\end{equation}
Here $x_i$ represents a condensate value on configuration $i$ 
and $x_{i + \Delta T}$ that on a configuration a further 
$\Delta T$ time units along in the ordered ensemble. $\Delta T = 1$ 
thus corresponds to adjacent configurations. 
Fig.~\ref{fig:autocorr} shows that nearby configurations in 
the ensembles are correlated and thus binning is necessary to 
obtain a reliable statistical error. A similar analysis applies
to masses and decay constants as discussed in~\cite{Dowdall:2011wh}. 

In the next section we describe the perturbative calculation of 
the condensate which we will then subtract from the raw results 
of Table~\ref{tab:valence}
to enable the nonperturbative condensate to be determined. 

\subsection{Perturbative calculation of $\langle \mathrm{Tr} M^{-1} \rangle$}
\label{subsec:pt}

\begin{center}
\begin{figure}[ht]
\includegraphics[width=0.3\hsize]{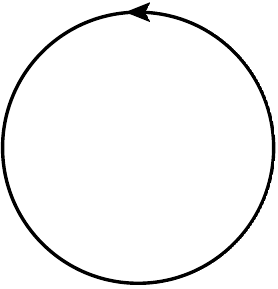}
\includegraphics[width=0.3\hsize]{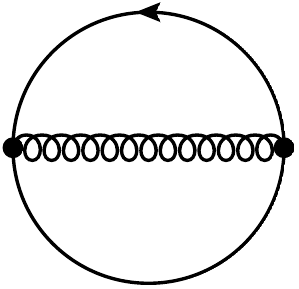}
\includegraphics[width=0.3\hsize]{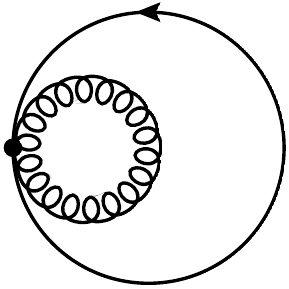}
\caption{Feynman diagrams for the calculation of the 
perturbative contribution to the quark condensate through $\mathcal{O}(\alpha_s)$.}
\label{fig:feyn}
\end{figure}
\end{center}

We computed the perturbative contribution $\langle \overline\psi\psi \rangle_{\rm PT}$ to the chiral condensate for the HISQ action through first-order in $\alpha_s$:
\begin{eqnarray}
   - a^3 \langle \overline\psi\psi \rangle_{\rm PT, HISQ} &=& am_0 \times \\
&& \left[ c_0(am_0) + c_1(am_0)\alpha_s + O(\alpha_s^2) \right]  \nonumber,
\label{PTcond}
\end{eqnarray}
where $am_0$ is the bare quark mass parameter that appears in the HISQ action. 
The Feynman diagrams required to this order are shown in Fig.~\ref{fig:feyn}. 
The perturbative quadratic ultraviolet divergence discussed in eq.(\ref{eq:contdiv}) shows up as finite values for the perturbative coefficients, as defined above, in the limit $am_0 \rightarrow 0$. 

\begin{center}
\begin{figure}[ht]
\includegraphics[width=0.9\hsize]{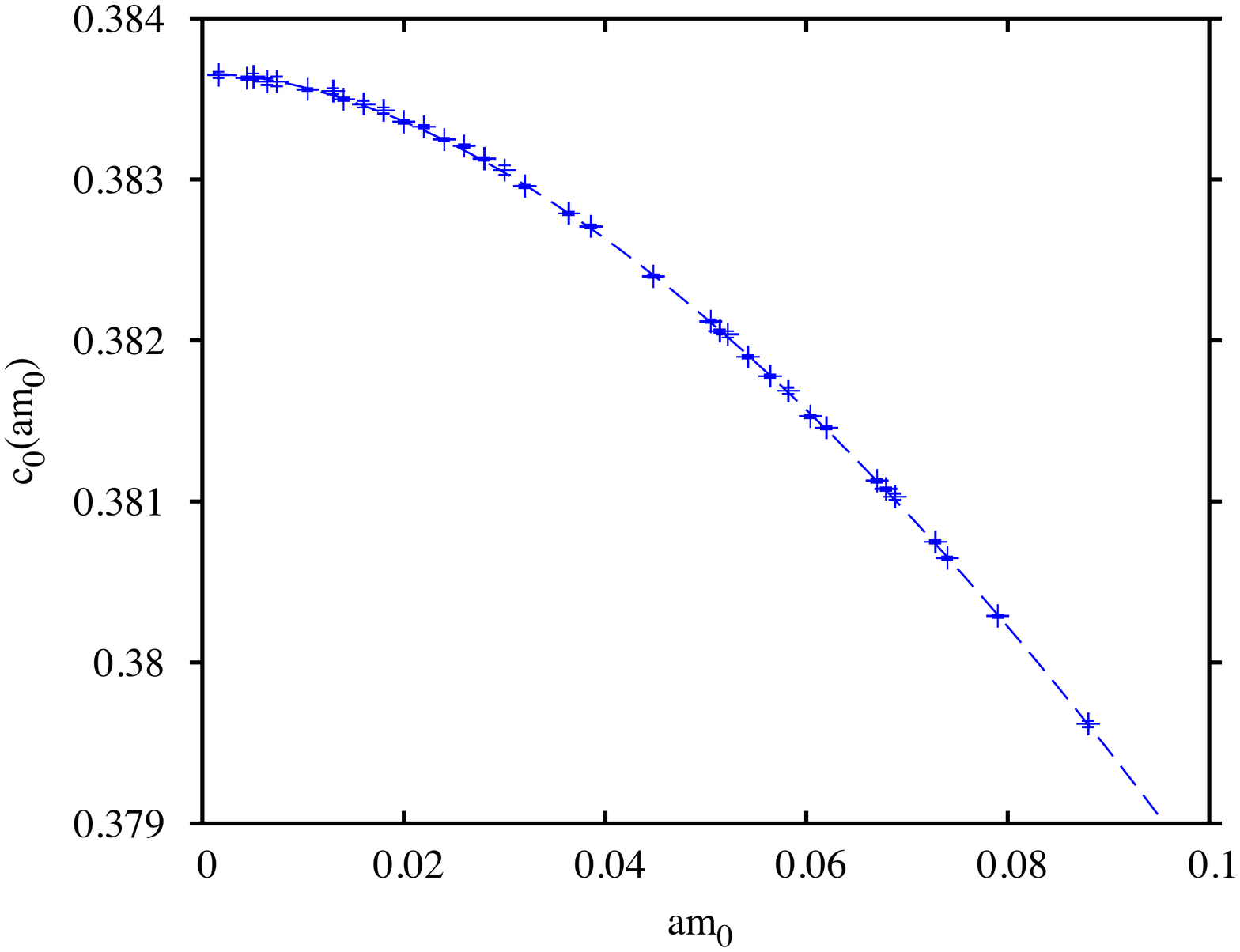}
\includegraphics[width=0.9\hsize]{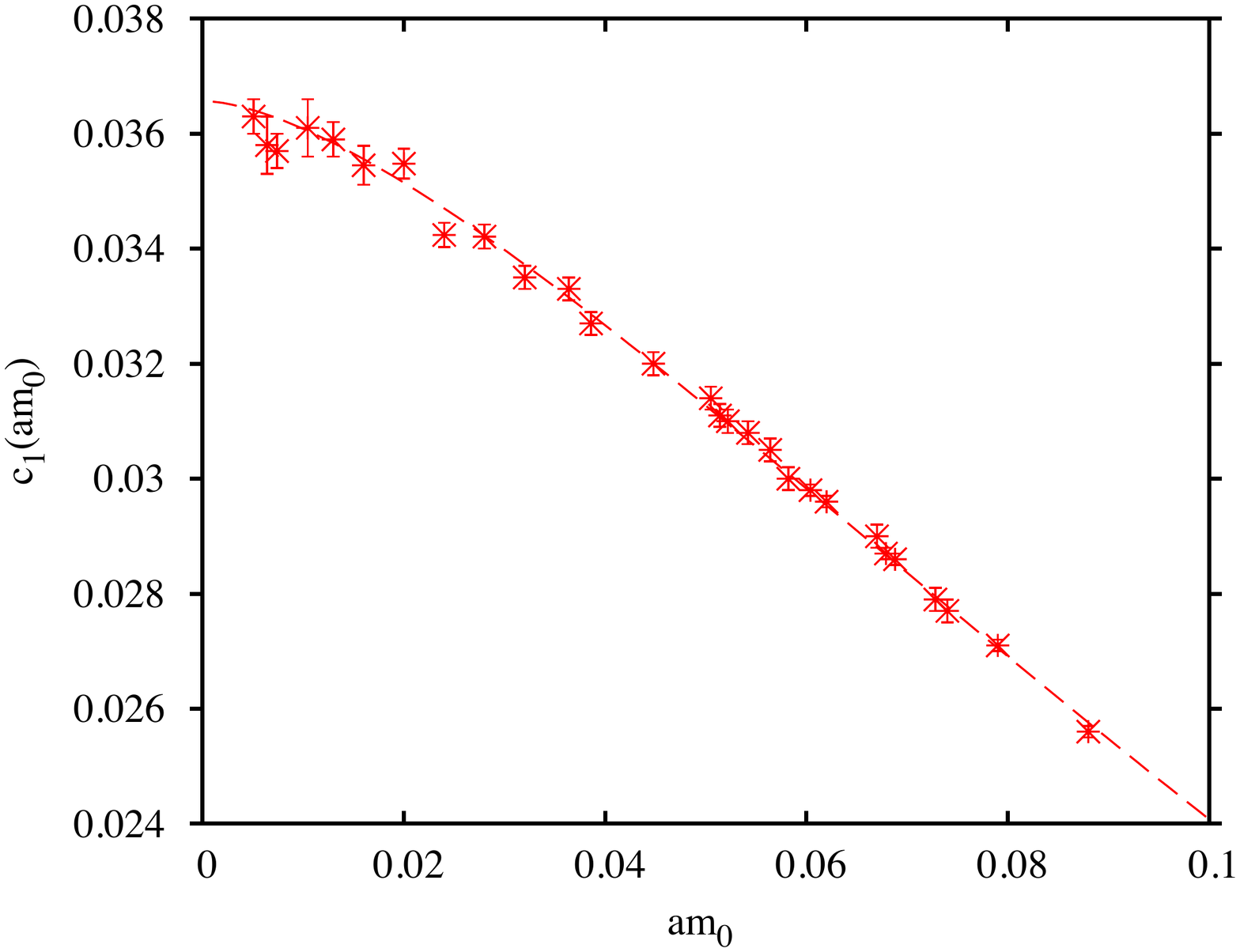}
\caption{\label{f:PTplot0}Zeroth- and first-order coefficients, $c_0$ and $c_1$ respectively, for the perturbative condensate, Eq.\ (\ref{PTcond}), versus the bare quark mass parameter $am_0$ in lattice units. The uncertainties in $c_0$ resulting from numerical evaluations of the lattice loop integral are not visible in that plot. The fits given in the text are plotted as dashed lines. }
\label{PTfigs}
\end{figure}
\end{center}

We computed the coefficients from numerical evaluation 
of the lattice loop integrals over a range of masses that 
includes the light and strange quark masses that we have used.  
A representative sample of our results is given 
in Table~\ref{PTtable}, and is illustrated in Fig.~\ref{PTfigs}.

An excellent fit to the perturbative coefficients in this range of small quark masses, $am_0 \alt 0.1$, can be obtained using the following parameterizations
\begin{equation}
   c_0(am_0) = c_{00} + (am_0)^2 \left[ c_{01}\log(am_0) + c_{02} \right] ,
\label{PTc0}
\end{equation}
and
\begin{align}
   c_1(am_0) & = c_{10} + (am_0)^2 \bigl[ c_{11}\log^2(am_0) \nonumber \\
                  & \quad + c_{12}\log(am_0) + c_{13} \bigr].
\label{PTc1}
\end{align}
Higher order terms in $am_0$ appear as discretisation errors 
in the comparison to $\overline{MS}$ to be done below and so can 
be ignored - they are negligible for the masses 
we are using in any case. 
The leading logarithm of $am_0$ at each order originates entirely from the 
infrared region of the loop momenta, and the respective 
coefficients $c_{01}$ and $c_{11}$ can easily be 
computed analytically. The values of these coefficients must and 
do agree with the values in the $\overline{MS}$ scheme~\cite{Braaten:1991qm}. 
At one-loop we also have a constraint on the sub-leading (single) 
logarithm of $am_0$ since, as discussed in Appendix~\ref{appendix:appc}, all $\log m$ 
terms must vanish in the difference between the vacuum 
expectation values of $m\overline{\psi}\psi$ in perturbation theory 
in the continuum and on the lattice.  
Allowing for the renormalisation between the $\overline{MS}$ mass 
and the HISQ bare mass: 
\begin{equation}
\overline{m}(\mu) = m_0 \left( 1 + \alpha_s \left[ -\frac{2}{\pi}\log a\mu + 0.1143(3) \right] + \ldots \right) ,
\label{eq:massren}
\end{equation}
we find that $c_{12}$ should have the value 0.2307(2). 
With the logarithmic terms fixed to their known values we can obtain 
the other coefficients in eqs.~(\ref{PTc0}) and~(\ref{PTc1}) from 
a fit to the values for $c_0(am_0)$ and $c_1(am_0)$ as a function of $am_0$. 
We find: 
\begin{align} 
   c_{00} & = 0.38366(1), \nonumber \\
   c_{01} & = 3/(2\pi^2) , \nonumber \\
   c_{02} & = -0.153(1) ,
\label{PTc0fit}
\end{align}
and
\begin{align}
   c_{10} & = 0.03657(7) , \nonumber \\
   c_{11} & = -6/\pi^3, \nonumber \\
   c_{12} & = 0.2307(2), \nonumber \\
   c_{13} & = 0.308(15) . 
\label{PTc1fit}
\end{align}
These fits are illustrated in Fig.~\ref{PTfigs} and 
reproduce our results for the coefficients to within their 
numerical integration errors, which are smaller than 
about $0.01\%$ for $c_0$, and $1\%$ for $c_1$.

\begin{table}[htb]
\begin{ruledtabular}
\caption{Zeroth- and first-order coefficients, $c_0$ and $c_1$ respectively, for the perturbative condensate, Eq.\ (\ref{PTcond}), for representative values of the bare quark mass parameter $am_0$ in lattice units. The uncertainties are from a numerical evaluation of the lattice perturbation theory loop integrals.}
\label{PTtable}
\begin{tabular}{lll}
\multicolumn{1}{c}{$am_0$} & \multicolumn{1}{c}{$c_0$} & \multicolumn{1}{c}{$c_1$} \\
\colrule
0.088      &    0.37962(2)    &    0.02561(12)  \\ 
0.079      &    0.38029(1)    &    0.02709(13)  \\
0.0728    &    0.38075(1)    &    0.02793(16)  \\
0.067      &    0.38113(1)    &    0.02902(18)  \\
0.062      &    0.38146(1)    &    0.02963(14)  \\
0.0564    &    0.38178(1)    &    0.03049(22)  \\
0.0505    &    0.38212(1)    &    0.03139(16)  \\
0.0448    &    0.38240(1)    &    0.03203(18)  \\
0.0386    &    0.38271(1)    &    0.03267(20)  \\
0.032       &    0.38296(1)    &    0.03349(24)  \\
0.028       &    0.38313(1)    &    0.03421(21)  \\
0.024       &    0.38325(1)    &    0.03424(21)  \\
0.020       &    0.38336(1)    &    0.03548(26)  \\
0.016       &    0.38347(2)    &    0.03545(34)  \\
0.01044  &    0.38356(1)    &    0.03614(49)  \\
0.00507  &    0.38364(2)    &    0.03631(31) 
\end{tabular}
\end{ruledtabular}
\end{table}

The perturbative determination of the vacuum expectation value of 
$\overline{\psi}\psi$ has also been done in the $\overline{MS}$ 
scheme, in~\cite{Braaten:1991qm}. 
The power divergence is missing in this case but, 
as discussed above, there are terms proportional to  $m^3\log m$. 
~\cite{Braaten:1991qm} finds:
\begin{eqnarray}
   &-& \langle \overline\psi\psi \rangle_{\rm PT, \overline{MS}}^{(\mu)} = \overline{m}^3(\mu) \times  \\
&& \left[ d_{01}l_m + d_{02} + \alpha_s\left(d_{11}l_m^2 +d_{12}l_m + d_{13}\right) + \ldots\right]  \nonumber
\label{PTmsbar}
\end{eqnarray}
where $l_m = \log (\overline{m}(\mu)/\mu)$ and
\begin{eqnarray} 
d_{01} &=& c_{01} = \frac{3}{2\pi^2} \nonumber \\
d_{02} &=& -\frac{3}{4\pi^2} \nonumber \\
d_{11} &=& c_{11} = -\frac{6}{\pi^3} \nonumber \\
d_{12} &=& \frac{5}{\pi^3} \nonumber \\
d_{13} &=& -\frac{5}{2\pi^3} 
\end{eqnarray}

As discussed in Appendix~\ref{appendix:appb} we must subtract the difference between 
the lattice QCD and $\overline{MS}$ perturbative calculations from 
our lattice QCD results to obtain the nonperturbative condensate 
in the $\overline{MS}$ scheme at the scale $\mu$. 
We work with the combination $m\overline{\psi}\psi$ 
which would be RG-invariant in the absence of this perturbative 
contribution and it is convenient to derive the subtraction needed 
in lattice units and as a function of the bare lattice quark 
mass. Using eq.~(\ref{eq:massren}) we obtain  
\begin{eqnarray}
\label{eq:deltapert}
\Delta_{\rm PT} &=& -a^4 \left( \langle m_0\overline{\psi}\psi \rangle_{\rm PT, HISQ} - \langle \overline{m}(\mu)\overline{\psi}\psi \rangle_{\rm PT, {\overline{MS}}} \right) \nonumber \\
&=&  c_{00}(am_0)^2 + \alpha_s c_{10}(am_0)^2   \\
&+& (am_0)^4 \left[ c_{01}l_{\mu} -0.077(1) \right] \nonumber \\
&+& \alpha_s (am_0)^4 \left[ c_{11} l_{\mu}^2 + 0.1340(2) l_{\mu} + 0.406(15) \right]   + \ldots ,\nonumber 
\end{eqnarray}
where $l_{\mu} = \log(\mu a)$.
This difference of perturbative expansions is now free of all 
logarithms of $m$ and therefore well-defined and infrared safe. 

\subsection{Determining the nonperturbative strange and light quark condensates}
\label{subsec:numbers}

\subsubsection{A first look at the results} 
\label{subsub:firstlook}
The physical 
condensate in the $\overline{MS}$ 
scheme at the scale $\mu$ is then defined by: 
\begin{equation}
\langle m \overline{\psi} \psi \rangle_{NP,\overline{MS}} (\mu) = a^{-4} \left( a^4\langle m\overline{\psi} \psi \rangle_0 - \Delta_{PT} \right) ,
\label{eq:ppbarnp}
\end{equation}
where $\langle m\overline{\psi}\psi \rangle_0$ is the numerical 
result from lattice QCD and $\Delta_{PT}$ is given 
through $\mathcal{O}(\alpha_s)$ 
as a function of the quark mass in Eq.~(\ref{eq:deltapert}). 
$\Delta_{PT}$ will also contain unknown higher order 
pieces in $\alpha_s$ that we can try to determine 
from a fit to the lattice QCD results. 
First we look at the effect of the calculated tree-level 
and one-loop contributions. 

\begin{figure*}
\includegraphics[width=0.4\hsize]{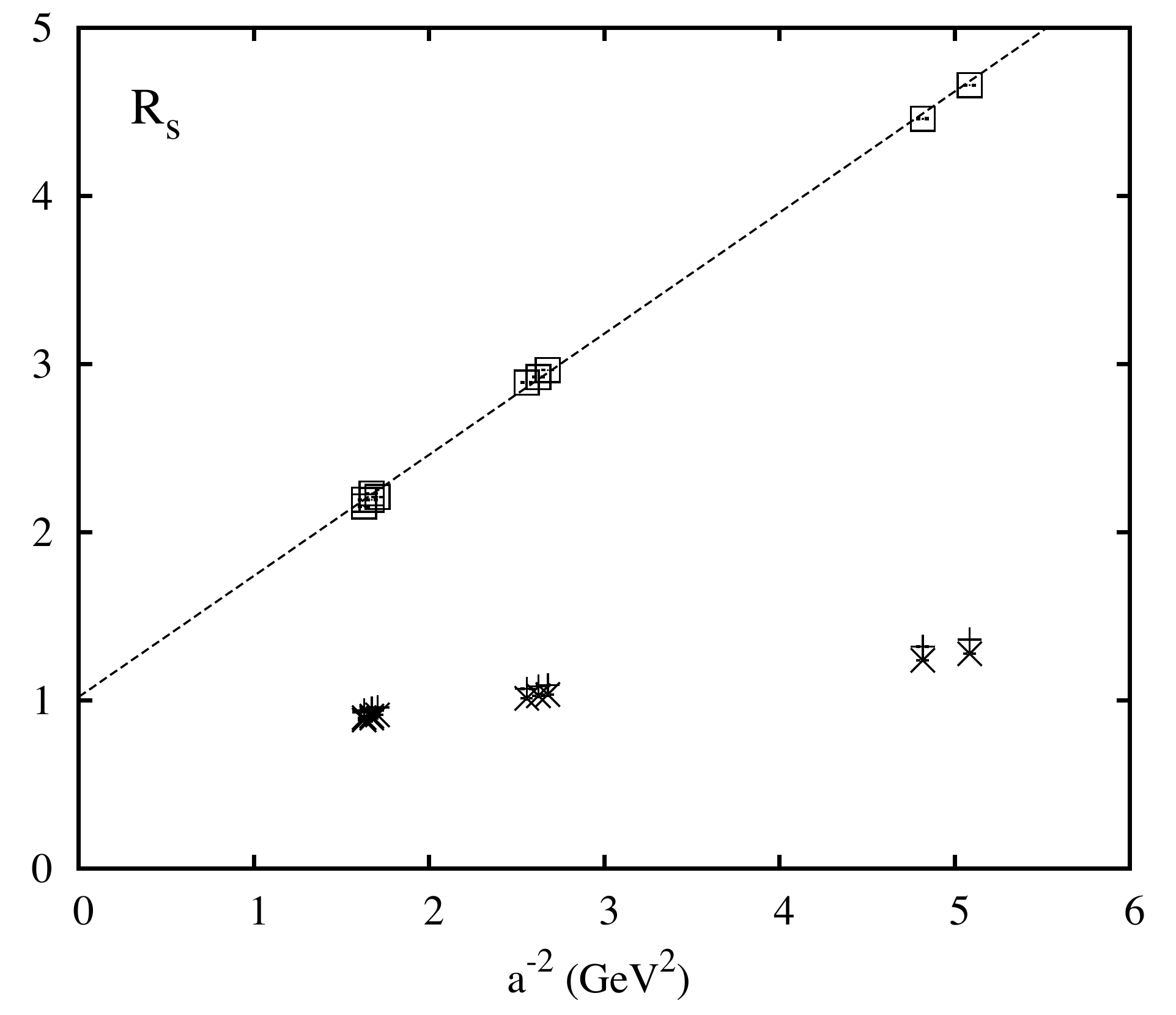}
\hspace{4em}
\includegraphics[width=0.4\hsize]{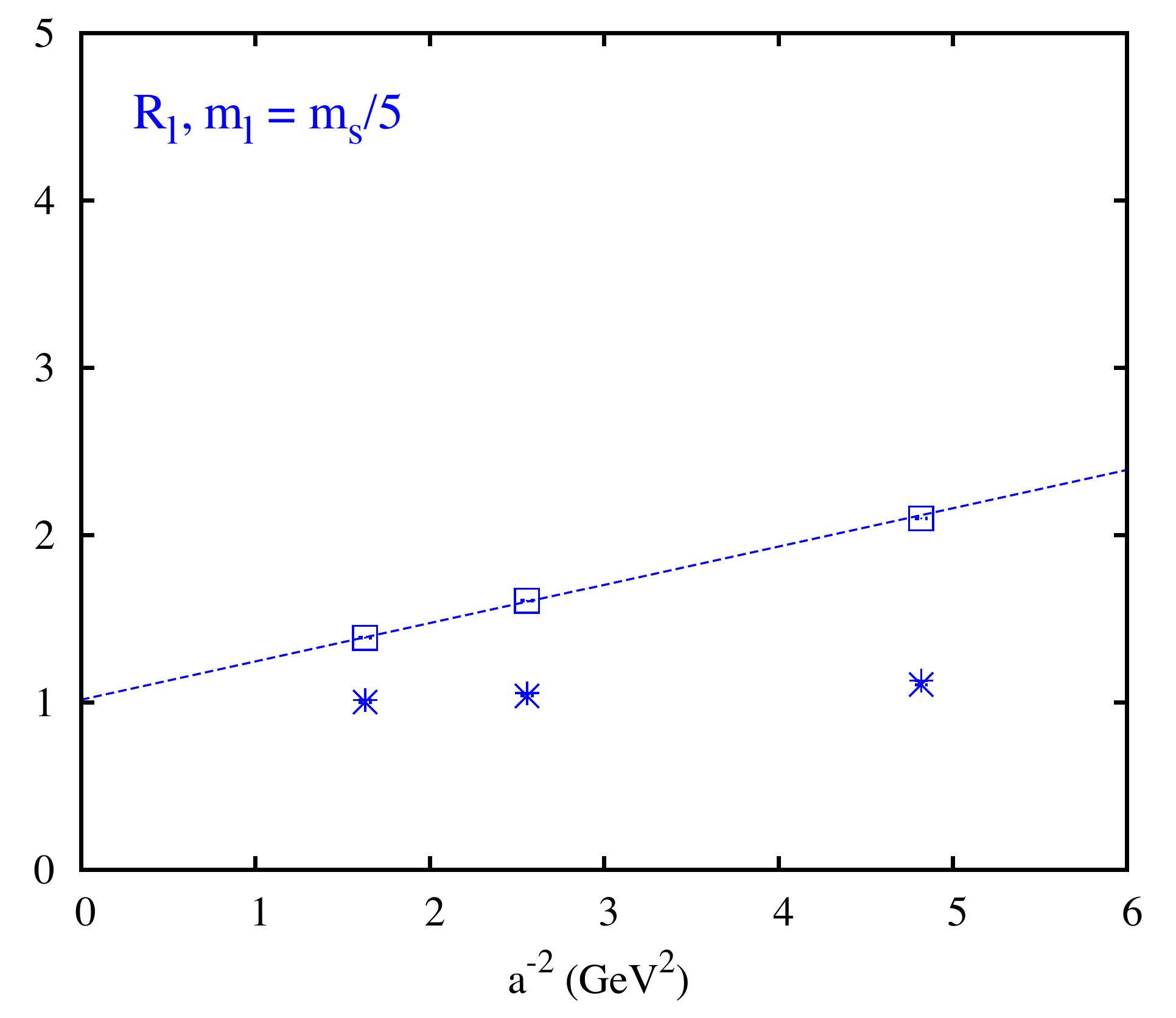}

\includegraphics[width=0.4\hsize]{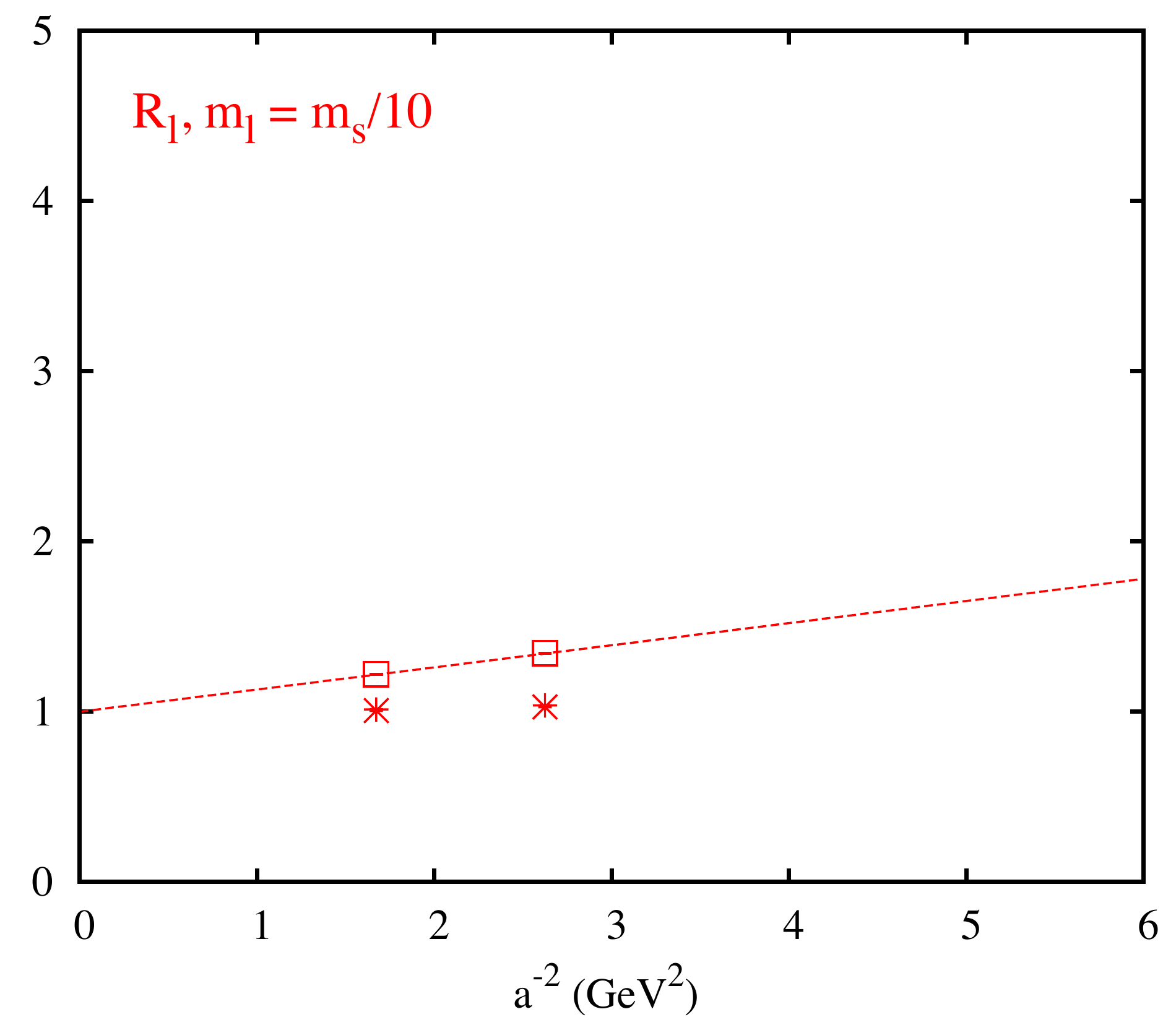}
\hspace{4em}
\includegraphics[width=0.4\hsize]{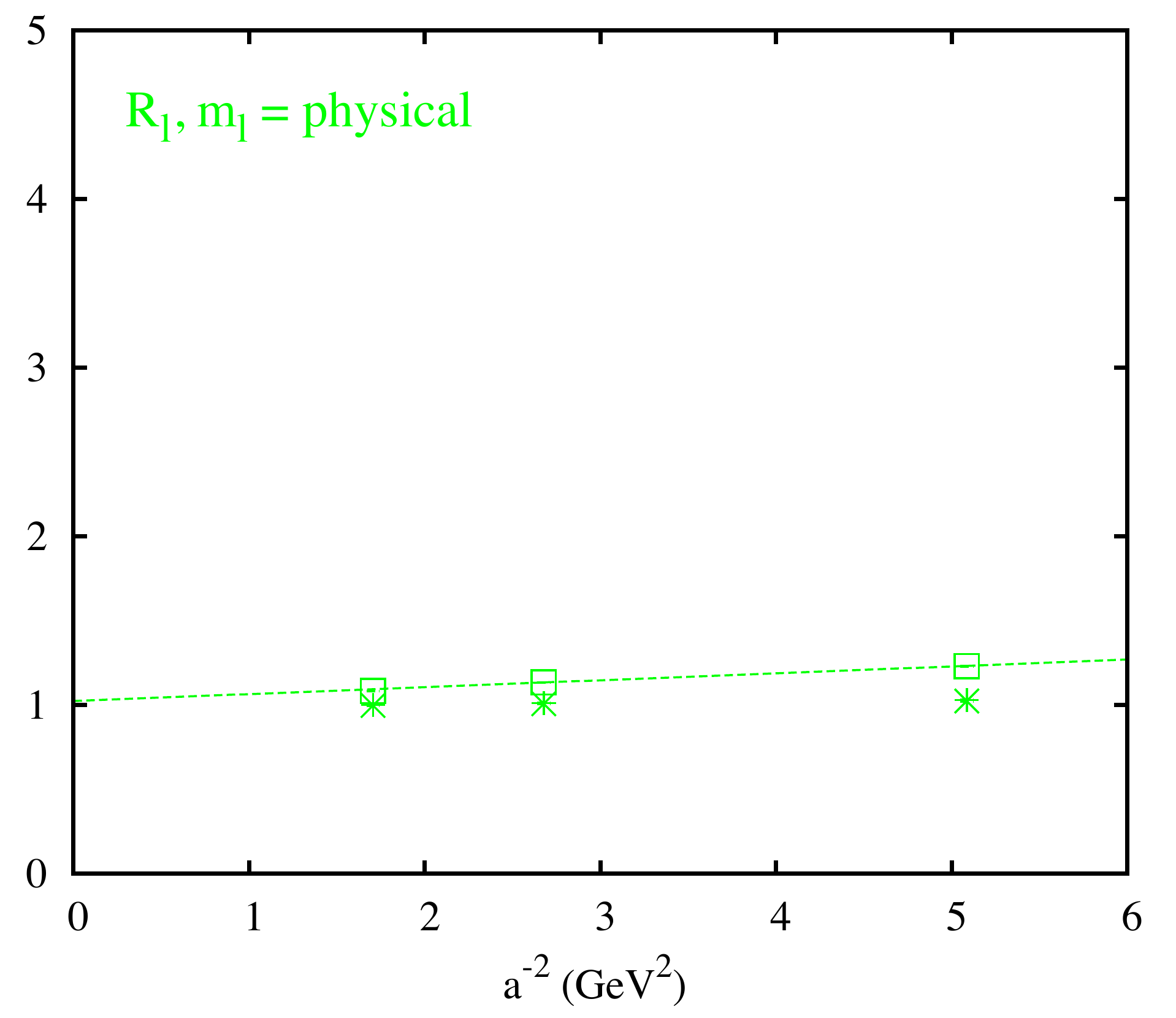}
\caption{
$R_q$, defined as the ratio of quark mass times condensate in the 
$\overline{MS}$ scheme at 2 GeV to the 
square of the meson mass times decay constant, as a function 
of the square of the inverse lattice spacing. Left 
to right and top to bottom shows strange quarks 
and light quarks with masses $m_s/5$, 
$m_s/10$ and the physical value. Squares use the 
unsubtracted condensate, pluses, the 
condensate after subtraction of the tree-level perturbative 
correction and crosses, the condensate after perturbative 
correction through one-loop. 
The value for $\alpha_s$ used to multiply the one-loop coefficient 
was $\alpha_V^{n_f=4}(2/a)$. 
Dashed lines illustrate very simple linear fits to the unsubtracted results 
as described in the text. 
}
\label{fig:unsub}
\end{figure*}

$\Delta_{PT}$ is a strong 
function of the quark mass in lattice units, 
dominated by the $(ma)^2$ terms that 
give rise to the quadrative divergence with inverse  
lattice spacing. This means that the relative 
size of the subtraction 
compared to the raw results varies strongly with quark mass 
and with lattice spacing, and this is reflected 
in the raw results before the subtraction is made. 
In Fig.~\ref{fig:unsub} the open squares show 
the unsubtracted results (i.e. setting $\Delta_{PT}$ to zero 
in Eq.~(\ref{eq:ppbarnp})) 
as a function of the square of the inverse 
lattice spacing for quarks at the four different masses that 
we have results for in Table~\ref{tab:valence}: strange quarks and light 
quarks of masses $m_s/5$ (sets 1, 4 and 7), $m_s/10$ (sets 2 and 5) 
and the physical value, $m_s/27$ (sets 3, 6 and 8).  

Instead of plotting the condensate results directly, 
the $y$-axis in Fig.~\ref{fig:unsub} is: 
\begin{equation}
R_l = -\frac{4m_l\langle \overline{\psi} \psi_l \rangle}{(f_{\pi}^2M_{\pi}^2)}
\label{eq:rldef}
\end{equation}
for light quarks and 
\begin{equation}
R_s = -\frac{4m_s\langle \overline{\psi} \psi_s \rangle}{(f_{\eta_s}^2M_{\eta_s}^2)} 
\label{eq:rsdef}
\end{equation}
for strange quarks. 
The values of the raw unsubtracted $R_q$ are determined 
directly from Table~\ref{tab:valence} 
using the $am$, $aM$, $af$ and $\langle \overline{\psi}\psi \rangle_0$ 
values given there. 

\begin{figure}
\includegraphics[width=0.9\hsize]{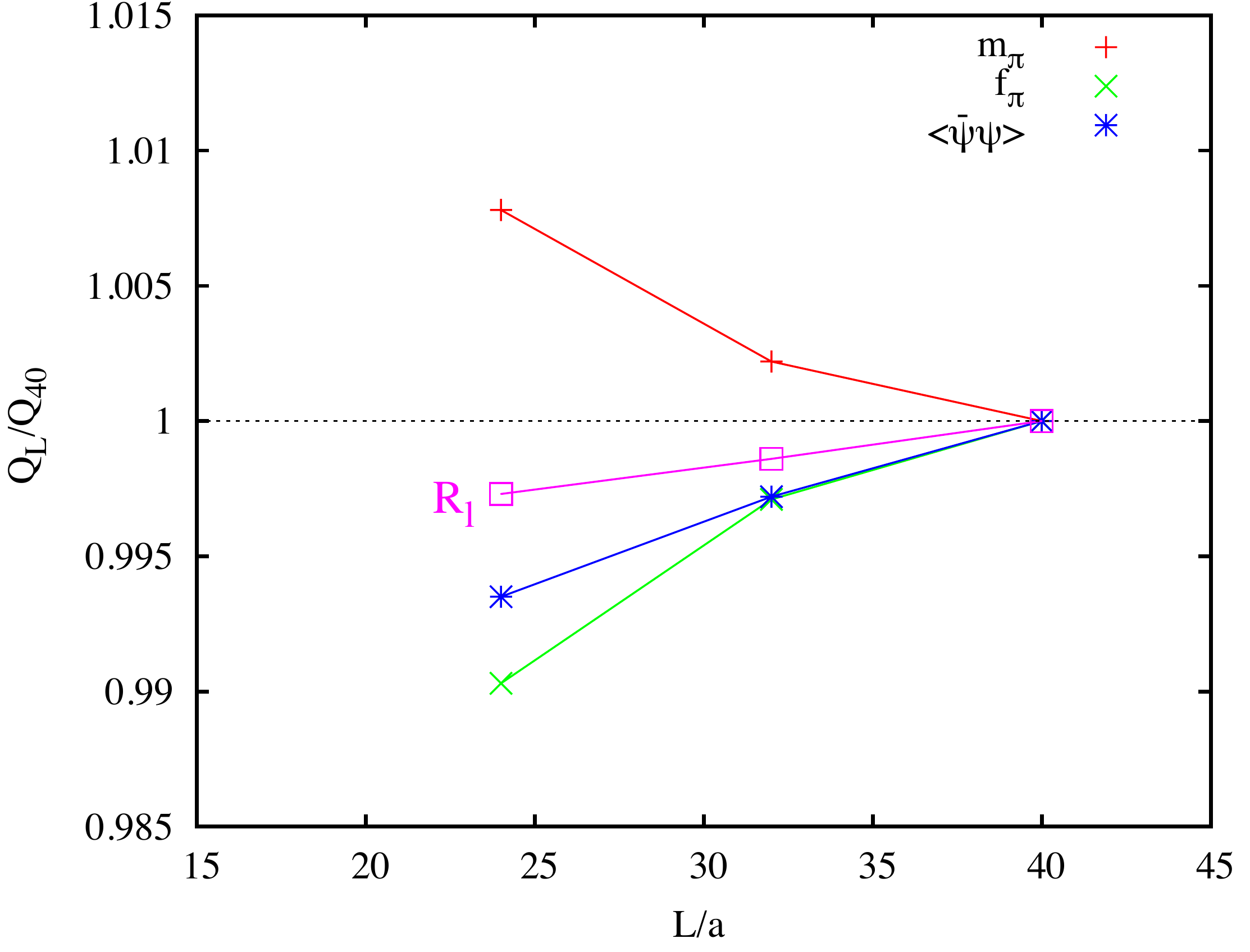}
\caption{
Finite volume effects in different quantities are 
illustrated by plotting the ratio of the quantity 
on lattices of spatial length, $L/a$, of 24 and 32 to that 
on lattices of spatial length 40. 
The lattices have the sea quark mass parameters
of coarse set 5. 
The quantities shown are the pion mass (red pluses), 
pion decay constant (green crosses) and 
unsubtracted light quark condensate (blue bursts). 
Pink squares give the result for the quantity $R_l$ 
defined in Eq.~(\ref{eq:rldef})~\cite{doug}.   
Statistical errors (not shown) are approximately 0.1\%. }
\label{fig:finitevol}
\end{figure}

The ratio $R_q$ is a good quantity to plot (and later to use 
in our fits) for a number of reasons: 
\begin{itemize}
\item{
$m\langle \overline{\psi} \psi \rangle$ is a physical 
renormalisation-group invariant 
quantity as $m \rightarrow 0$, up to discretisation errors, 
as is clear from the GMOR relation.  
The division by the square of the meson decay constant 
times its mass makes a dimensionless 
ratio which is convenient but it is also one that 
(from the GMOR relation) we expect 
to be close to 1. }
\item{ Using the ratio $R_q$ also reduces 
the effect of any slight mistuning of 
quark masses since the quark mass multiplied in the 
numerator cancels against the square of the meson mass in the denominator. 
The tuning of $m_s$ uses the $\eta_s$ decay constant, 
as described in~\cite{Dowdall:2011wh}. This means that, 
by definition, $f_{\eta_s}$ does not contain 
discretisation errors that would mask the identification 
of the pieces that diverge as $a \rightarrow 0$. } 
\item{ Finally the ratio has reduced 
finite-volume effects over that in either 
the numerator or denominator. This is expected 
from the fact that chiral loop effects, which
are sensitive to the volume, cancel 
in $R_q$~\cite{Gasser:1984gg, Jamin:2002ev}. 
This is illustrated in 
Fig.~\ref{fig:finitevol} where we show results~\cite{doug} for 
pion mass, decay constant and (unsubtracted) light quark 
condensate as well as the ratio $R_l$ of Eq.~(\ref{eq:rldef}) 
for ensembles with $am_{l,sea}=0.00507$ and $am_{s,sea}=0.0507$ 
and three different spatial volumes. The spatial volumes 
correspond to a spatial length in lattice units of 24, 32 and 40. 
The set with $L/a = 32$ is our set 4 (see Table~\ref{tab:params}).    
For each quantity we plot the ratio of the value at 
$L/a$ to that at $L/a=40$. It is clear from the plot that 
the finite volume dependence in each of $m_{\pi}$, $f_{\pi}$ 
and $\langle \overline{\psi}\psi \rangle_l$ is cancelled 
to a very high level of accuracy (0.1(1)\% for set 4) in $R_l$. }
\end{itemize}

Figure~\ref{fig:unsub} shows clearly 
the presence of a quadratic divergence 
with $a^{-2}$ in the raw results. 
This is very `clean' 
in our calculations because the form of the divergence is very constrained. 
Only a term of the form $m_q/a^2$ is allowed in 
$\langle \overline{\psi} \psi \rangle$ for staggered quarks, i.e. 
no term of the form $m_q^2/a$ can appear. In the ratio $R_q$ 
this term takes the form $Cm_q^2/a^2$ where $C$ depends on the 
meson mass and decay constant. The HISQ formalism has very 
small discretisation errors, as is clear from the decay constant and 
meson mass results in~\cite{Dowdall:2011wh}, and so there is little additional 
$a$-dependence to confuse the analysis of the divergent pieces. 

Because the power divergence is so dominant 
it is tempting to try to fit the unsubtracted results for 
$R_s$ to a very simple form: $A+B/a^2$. 
This is in fact possible (it is important to include 
the error in the inverse lattice spacing when doing this 
since this is larger than the error in $R_q$) and 
we obtain $1.02(3)+0.725(3)/a^2$ which 
is the dashed line in lefthand plot of Fig.~\ref{fig:unsub}. 
We also obtain $1.015(11) + 0.229(5)/a^2$ for $R_l$ with $m_l=m_s/5$ 
and $1.00(1) + 0.130(6)/a^2$ for $R_l$ with $m_l = m_s/10$, 
shown in the next two plots in Fig.~\ref{fig:unsub}. 
These fits are too naive to be useful, 
as we shall see below, because they miss out many 
important terms. Consequently the 
value and error of the intercept, $A$, is 
unreliable for extracting a nonperturbative 
result for $R_q$, especially in the $s$ quark case. 
However, the fits do illustrate that the ratio of 
slopes is that expected for a term 
that behaves as $m_q^2/a^2$ (although the simple fit does not 
allow for the running of the lattice bare quark mass with 
scale). The ratio 
of slopes between that for $s$ and for 
$l$ with $m_l=m_s/5$ is 3.2 which corresponds 
approximately to 5 (for the ratio of one power of the quark 
masses when the other power is cancelled by 
the square of the meson mass) divided by 
the ratio of the square of the decay constants
from Table~\ref{tab:valence}.

Figure~\ref{fig:unsub} compares 
results for $R_q$ in which the tree-level piece of $\Delta_{PT}$
has been subtracted from the raw values of 
$m\langle \overline{\psi}\psi \rangle$ 
following Eq.~\ref{eq:ppbarnp}. We take
$\mu$ to be 2 GeV. 
These results are indicated by pluses. 
Now the slope in $a^{-2}$ is much 
smaller since the most of the divergence has been removed. 
This makes the results more sensitive to the 
form of the remaining pieces of the divergence 
and the simple linear fits that were made to the 
unsubtracted data are no longer possible. 

The $\mathcal{O}(\alpha_s)$ perturbative contribution is very small 
for HISQ quarks and makes very little difference to the perturbative 
subtraction. 
The crosses show the results taking $\Delta_{PT}$ to be the 
full calculated perturbative subtraction through $\mathcal{O}(\alpha_s)$ 
given in Eq.~\ref{eq:deltapert}. 
We have used $\alpha_V^{n_f=4}(2/a)$~\cite{McNeile:2010ji} for 
the $\alpha_s$ value multiplying the one-loop coefficient, but the 
coefficient itself is so small that variations in scale for $\alpha_s$ 
make no difference. 
The crosses are barely distinguishable from 
the pluses giving the tree-level subtracted numbers.  

It is clear from Fig.~\ref{fig:unsub} that 
there is still some divergence in $a^{-2}$ 
left in $R_q$ after subtraction of the perturbative
contribution through one-loop. 
This is not surprising since we know that 
$\Delta_{PT}$ will have higher order terms 
in $\alpha_s$. 
The challenge now is to fit the one-loop 
subtracted $R_q$ allowing for these higher 
order terms  and thereby obtain the physical, 
nonperturbative, results for the strange and 
light quark condensates. We will fit both 
$R_s$ and $R_l$ simultaneously and use the 
known mass dependence of the unknown higher order terms 
to constrain them. 
At the same time we will allow for higher-order 
non-divergent mass-dependent terms from perturbation theory 
as well as physical, non-perturbative, dependence on 
the quark mass. 
Possible dependence on positive powers of $a$, i.e. 
discretisation errors, must also be included.   

\subsubsection{Determining a physical result from fitting} 
\label{subsub:fits}
We now describe the full fit to the results that we use to 
determine the final physical values for $\langle \overline{s}s \rangle$ 
($\equiv \langle \overline{\psi}\psi_s \rangle$) 
and $\langle \overline{l}l \rangle$ 
($\equiv \langle \overline{\psi}\psi_l \rangle$) 
at the 
physical strange and light quark masses (where $m_l = (m_u+m_d)/2$). 
We take the following form for the ratio $R_q$: 
\begin{equation}
R_{q, 0}(a, am_q) = R^{(q)}_{\mathrm{NP, phys}} + \delta R_{PT} + \delta R_{a^2} + \delta R_{\chi} + \delta R_{\mathrm{vol}}.
\label{eq:fit}
\end{equation}
$R_{q,0}$ are the raw results obtained from Table~\ref{tab:valence}, 
$R_{\mathrm{NP, phys}}$ is the final physical result in the 
$\overline{MS}$ scheme at 2 GeV. The $\delta R$ 
terms represent fitted or known dependence on $a$ and $am_q$. 
We use Bayesian techniques~\cite{gplbayes} to perform the fits so that 
we can add many higher order terms as part of each $\delta R$ 
with constrained coefficients. 
This makes sure that the final error on $R_{\mathrm{NP, phys}}$ 
is not underestimated by ignoring the existence of higher order 
corrections.  

$\delta R_{PT}$ contains the known tree-level and one-loop perturbative 
results given in section~\ref{subsec:pt}. In addition we include unknown 
higher-order terms. For the $a^{-2}$ divergence these take the form: 
\begin{equation}
\delta R_{PT, div} = a_n \frac{4 \alpha_s^n (am_q)^2}{(af_{\pi})^2(aM_{\pi})^2} 
\label{eq:rptdiv}
\end{equation}
with the analogous term for the strange quark case, with the same $a_n$. 
$a_n$ is a coefficient 
whose prior we take to be $0.0 \pm 4.0$ and we allow for $n = 2$, 3 and 4. 
Note that a prior width of 4.0 is conservative given the size 
of the corresponding coefficient at tree-level and one-loop. 
For the non-divergent pieces we take: 
\begin{equation}
\delta R_{PT,non-div} = c_n \frac{4 \alpha_s^n (am_q)^4}{(af_{\pi})^2(aM_{\pi})^2}
\label{eq:rptnondiv}
\end{equation}
again with the analogous term for the strange quark case, with the 
same coefficient $c_{n}$. We take $n = 2$, 3 and 4.  
$c_n$ in principle contains a sum of powers of $\log(a\mu)$ up 
to $\log^{n+1}(a\mu)$. However, since $\log(a\mu)$ is small 
for $\mu = 2 \mathrm{GeV}$ and our range of lattice spacings, 
these pieces are negligible and do not affect the fit 
and we simply take a prior on $c_n$ of 0.0(4.0).
Again this is conservative given the results at tree-level and one-loop. 
For $\alpha_s$ we use $\alpha_V^{n_f=4}(2/a)$~\cite{McNeile:2010ji} and discuss below 
the dependence of the results on changing $2/a$ to a different scale. 
$\alpha_V^{n_f=4}(2/a)$ takes values from 0.35 on the 
very coarse lattices to 0.26 on the fine ensembles.

$\delta R_{a^2}$ allows for discretisation errors. We take the form 
\begin{equation}
\delta R_{a^2} = \sum_{i=1}^2 d_i \left(\frac{\Lambda a}{\pi}\right)^{2i} .
\end{equation}
Only even powers of $a$ appear in discretisation errors 
for staggered quarks and we take their scale to be 
set by $\Lambda \approx$ 1 GeV. 
Since all tree level errors at $\mathcal{O}(a^2)$ 
are removed in the HISQ formalism we take the prior for 
$d_1$ to be $\mathcal{O}(\alpha_s)$ i.e. 0.0(0.3). Higher 
order $d_i$ are given the prior 0.0(1.0). We include $a^2$ and $a^4$ 
terms, but have checked that higher order terms have very little effect. 
In addition we include a mass-dependent discretisation error in the form 
$e(am)^2$, giving $e$ a prior of 0.0(1.0). This allows for a 
number of effects, one of which could be mixing with a gluon condensate. 
This has negligible impact. 

$\delta R_{\chi}$ includes the valence and sea quark mass dependence 
that allows us to extrapolate to physical light quark masses and 
interpolate between the strange quark masses that we have to the 
physical strange quark mass. The chiral corrections to the GMOR 
relation were analysed in~\cite{Gasser:1984gg} (see also~\cite{Jamin:2002ev}). 
The leading corrections are particularly simple because the 
chiral logarithms cancel to leave a correction proportional 
to $M_{\pi}^2$. We allow for both $M_{\pi}^2$ and $M_{\pi}^4$ 
terms in our light quark mass fits by defining a chiral expansion parameter 
\begin{equation}
x_l = \frac{M_{\pi}^2}{2(\Lambda_{\chi})^2}, 
\label{eq:chiral}
\end{equation}
with $\Lambda_{\chi}=$ 1.0 GeV,
and taking 
\begin{equation}
\delta R_{\chi,val} = \sum_{i=1}^2 g_i^{(l)} x_l^i .
\label{eq:rchi}
\end{equation}
We fit the $m_q = m_s/5$, $m_s/10$ and $m_{l,phys}$ results with
this form taking the prior on the $g_i$ coefficients to 
be 0.0(2.0). This allows for a linear term of approximately 
the size expected in~\cite{Jamin:2002ev}. Higher order terms than $x_l^2$ 
have no effect. 

The chiral expansion of Eq.~(\ref{eq:rchi}) combines with 
the $R_{\mathrm{NP,phys}}$ parameter in Eq.~\ref{eq:fit} 
to define the physical non-perturbative light condensate 
with its mass dependence. Since the GMOR relation is 
exact as $m_q \rightarrow 0$ we enforce this by taking the prior on 
$R^{(l)}_{\mathrm{NP,phys}}$ to be 1.0000(5). 
Differences from 1 because of residual lattice finite volume 
effects up to 0.5\% are allowed for as described in 
the paragraph above. 

The data for $R_s$ is fit simultaneously with that for 
the light quarks, because they share parameters for 
the perturbative subtraction. However, we largely decouple 
the physical parameters  because the strange quark 
is relatively far from the chiral limit and we would 
need a lot of parameters for a chiral expansion to 
connect light and strange quarks. Instead we 
allow a separate parameter for $R^{(s)}_{\mathrm{NP,phys}}$ with 
the broad prior of 1.0(5). We take the same form for the valence mass 
dependence as in Eq.~(\ref{eq:rchi}) where now 
\begin{equation}
x_s = \frac{(M_{\eta_s}^2-(0.6893(12))^2)}{2(\Lambda_{\chi})^2}, 
\label{eq:schiral}
\end{equation}
but this now simply allows for slight mistuning of the 
strange quark on some ensembles, and the fact that we 
have two values for the strange quark mass on sets 1 and 2. 
0.6893 is the physical value for the $\eta_s$ mass 
determined in~\cite{Dowdall:2011wh}. 
The priors on $g_i^{(s)}$ are the same as those on 
$g_i^{(l)}$. Finite volume effects are expected to 
be completely negligible for $R_s$ because they 
are negligible for the components of $R_s$. 

The strange quark results in Fig.~\ref{fig:unsub} show 
some very small sensitivity to the sea light quark masses 
if we compare results on sets 1 and 2 and sets 3 and 4.  
We therefore allow an additional linear dependence on the light 
quark mass in the sea in $\delta R_{\chi}$ of the form 
\begin{equation}
\delta R_{\chi,sea} = \sum_{i=1}^2 k_i\left(\frac{\delta m_{sea}}{10m_{s,phys}}\right)^i
\label{eq:rchisea}
\end{equation}
where 
\begin{equation}
\delta m_{sea}= (2m_{l,sea}+m_{s,sea})-(2m_{l,phys}+m_{s,phys}).
\label{eq:delta}
\end{equation}
We take $m_{l,phys} = m_s/27.4$~\cite{milcreview} and 
$m_{s,phys}$ values determined from $M_{\eta_s}$ as 
in~\cite{Dowdall:2011wh}. 
The prior for coefficient $k_i$ is taken as 0.0(1.0) which 
is conservative given the small effects observed in 
the results. We take $\delta R_{\chi,sea}$ to be common 
to both $R_s$ and $R_q$ for the light quarks. 
We note here also that the absence of chiral loop effects 
at this order means that staggered quark taste-changing effects are 
also absent. They can be handled, if necessary, with a 
sea-quark mass dependent $a^2$ term~\cite{Dowdall:2011wh}. Including 
such a term here makes no difference to the physical result. 

$\delta R_{\mathrm{vol}}$ allows for remaining 
finite volume effects. These are small, 
as demonstrated in Fig.~\ref{fig:finitevol}.
They do produce a small systematic effect, however, 
because the lattice size in units of the pion 
mass, $M_{\pi}L$, is somewhat smaller on the lattices
with smallest $m_{u/d}$.  
We take:
\begin{equation}
\delta R_{\mathrm{vol}} = v e^{-ML}
\label{eq:fvol}
\end{equation}
where $M$ is the pseudoscalar meson mass made of that quark 
($M_{\pi}$ for $R_l$ and $M_{\eta_s}$ for $R_s$) and 
$L$ is the linear extent of the lattice from Table~\ref{tab:params}. 
Coefficient $v$ is taken to have prior 0.0(0.2), 
consistent with Fig.~\ref{fig:finitevol}. 

Fitting $R_s$ and $R_q$, $m_q=m_s/5$, $m_s/10$ and $m_{l,phys}$, results 
simultaneously to the form in Eq.~(\ref{eq:fit}) readily 
produces good fits with $\chi^2/\mathrm{dof} \approx 0.8$ for 
18 degrees of freedom. The final fitted result for $R_{u/d}$ 
and $R_s$ (evaluated from $R^{(q)}_{\mathrm{NP,phys}}$ and 
$\delta R_{\chi}$ taken at the appropriate physical masses) 
is robust to the addition of higher order terms in 
the various corrections. 

We take our final results from 
using $2/a$ for the scale of $\alpha_s$. The results do not 
change significantly as this is varied (although the fitted 
coefficients $a_n$ do change). 
Our fits 
return a substantial value for the coefficient of the power divergent term 
at $\mathcal{O}(\alpha_s^2)$, $a_2$, of around 2.0, for $d/a=2/a$. 
This is substantially larger than 
that seen at one-loop but not a particularly large value 
for a perturbative coefficient in general. It would simply imply that 
the small coefficient at one-loop for the HISQ action 
is not repeated at higher orders.  
We also find that the chiral correction to the GMOR for light quarks is 
substantial and negative ($g_1^{(l)}=-1.7(6)$). 
This will be discussed further below. 

\begin{center}
\begin{figure}[ht]
\includegraphics[width=0.9\hsize]{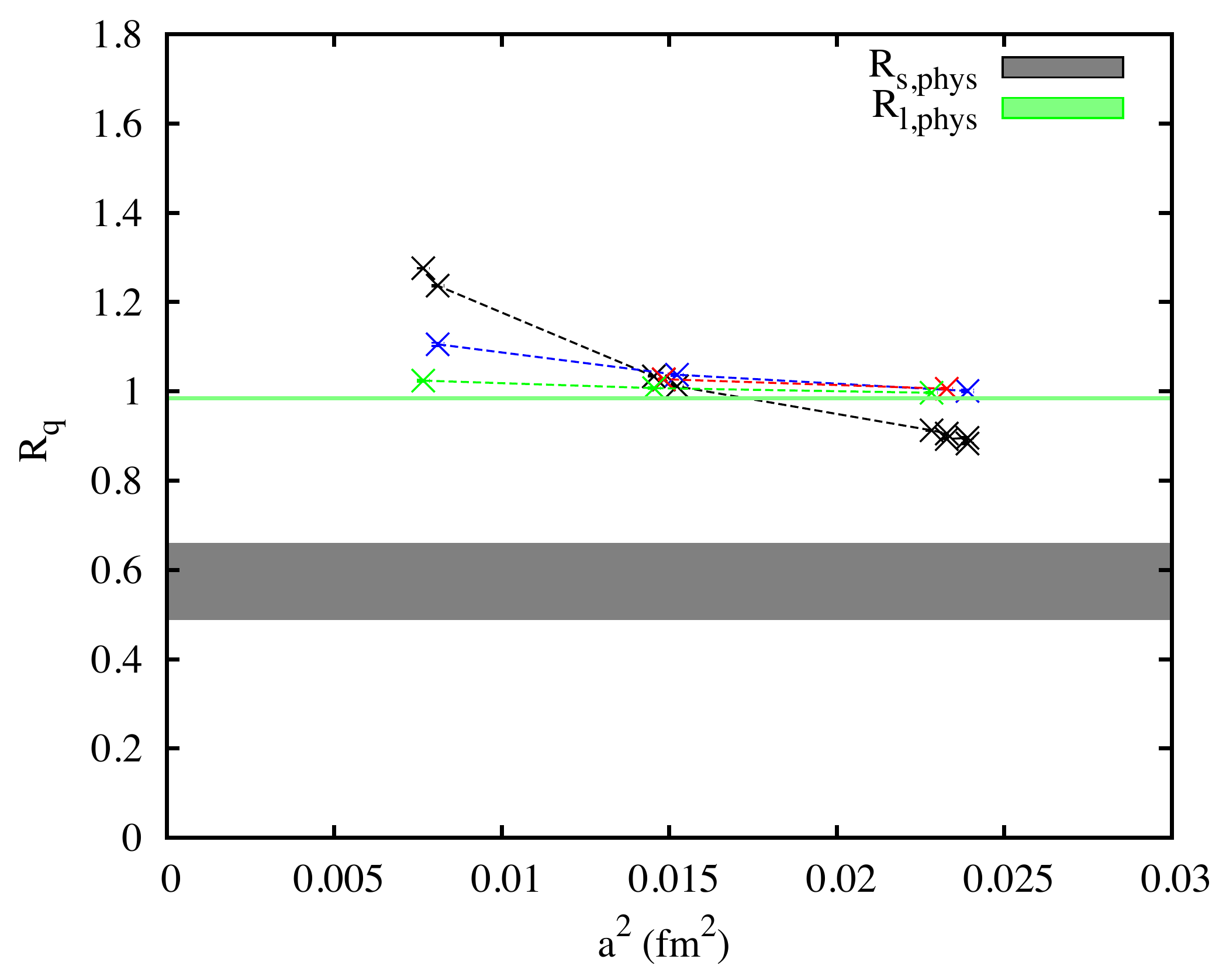}
\caption{Results from fitting the ratio $R$ 
for three different quark masses as described 
in the text. The crosses show the lattice 
QCD results after subtracting the perturbative 
values through $\mathcal{O}(\alpha_s)$. Black is 
for $s$ quarks, blue for quarks with mass $m_s/5$, 
red for quarks with mass $m_s/10$ and green for 
quarks at the physical light quark mass. 
The dashed lines simply join the points of 
matching color for clarity. 
The filled bands show the physical curves for 
strange (black) and light (green) quarks, once the full subtraction 
of the fitted perturbative contribution is made 
and masses are set to their physical values. 
The bands include the full error from the fit. 
}
\label{fig:rhisqfit}
\end{figure}
\end{center}

The fit results are shown in Fig.~\ref{fig:rhisqfit}. 
The data points (crosses) correspond to the lattice 
QCD results after subtraction of the perturbative 
contribution through $\mathcal{O}(\alpha_s)$ (as in Fig.~\ref{fig:unsub}). 
The filled bands show the fitted curves when the full 
fitted perturbative contribution ($\delta R_{PT}$) is 
subtracted and masses and decay constants are set to the 
physical values corresponding to the $s$ quark 
and the light quark (for this we use $M_{\pi} = M_{\pi^0}$).  
These bands include the full error from the fit. 

Our final physical results for $R_q$ are the 
key results from this paper. 
\begin{eqnarray}
R_{l,phys} &=& -\frac{4m_l\langle \overline{\psi} \psi_l \rangle_{\overline{MS}}(2 \mathrm{GeV})} {(f_{\pi}^2M_{\pi}^2)} \nonumber \\
R_{s,phys} &=& -\frac{4m_s\langle \overline{\psi} \psi_s \rangle_{\overline{MS}}(2 \mathrm{GeV})}{(f_{\eta_s}^2M_{\eta_s}^2)} 
\label{eq:rphysdef}
\end{eqnarray}
We find:
\begin{eqnarray}
R_{s,phys} &=& 0.574(86) \nonumber \\
R_{l,phys} &=& 0.985(5) \nonumber \\
\frac{R_{s,phys}}{R_{l,phys}} &=& 0.583(84) .  
\label{eq:rval}
\end{eqnarray} 

\begin{table}
\caption{
Error budget for the quantities $R_{s,phys}$, $R_{l,phys}$ and their 
ratio defined in the text. Errors are given as percentages 
of the final physical result. 
}
\label{tab:errbudget}
\begin{ruledtabular}
\begin{tabular}{llll}
 & $R_{s,phys}$ & $R_{l,phys}$ & $\frac{R_{s,phys}}{R_{l,phys}}$ \\
\hline
statistics & 6.1 & 0.2 & 5.1 \\ 
lattice spacing & 10.0 & 0.3 & 9.7 \\
finite volume & 1.5 & 0.03 & 1.5 \\
$\alpha_s$ value & 1.7 & 0.06 & 1.7 \\
fitting power divergence & 7.5 & 0.3 & 7.2   \\
other perturbative subtraction & 1.3 & 0.07 & 1.3    \\
$\chi$al extrap./interp. ($m_s$)  & 3.0 & 0.1 & 2.9 \\
$\chi$al extrap./interp. ($m_l$)  & 4.5 & 0.2 & 4.3 \\
$a \rightarrow 0$ extrap. & 1.9 & 0.05 & 1.9 \\
sea mass effects & 0.5 & 0.01 & 0.5 \\
\hline
Total & 15  & 0.5  & 14.5  \\
\end{tabular}
\end{ruledtabular}
\end{table}

The complete error budgets for $R_{s,phys}$, $R_{l,phys}$ and their 
ratio are given in Table~\ref{tab:errbudget}. 
The substantial 15\% error that we have in $R_{s,phys}$ 
reflects the difficulty of extracting 
a physical result from a power divergent quantity.
For $R_l$ the error is 17 times better largely because the 
slope of the divergent piece is 15 times smaller. 
Errors in $R_{s,phys}$ are dominated 
by errors from the lattice spacing and from fitting the 
remaining power divergent subtraction terms. 
There are also substantial errors from statistics 
and from tuning to the light and strange physical 
mass points. This is done by tuning the appropriate 
meson masses through the term $\delta R_{\chi,val}$ 
in Eq.~\ref{eq:rchi}. This term depends on the lattice 
spacing through the definition of $x_l$ (Eq.~\ref{eq:chiral}) 
and $x_s$ (Eq.~\ref{eq:schiral}), because the meson 
masses appear in GeV units in these terms. The 
uncertainties in these terms then becomes correlated 
with the fit to the power divergence, increasing the 
uncertainty.   
For $R_l$ the power divergence is much less of an issue,  
but these same terms dominate the final error there as 
well. 

The results for $R_l$ and $R_s$ can be converted 
to values of the condensate using the lattice 
result for $f_{\eta_s}$ = 0.1819(5) GeV and 
$M_{\eta_s}$ = 0.6893(12) GeV~\cite{Dowdall:2011wh}, 
and experimental values for $f_{\pi}$ (0.1304(2) GeV) and $M_{\pi}$ 
(0.13498 GeV). 
We obtain: 
\begin{eqnarray}
m_s\langle \overline{\psi}\psi \rangle_s^{\overline{MS}}(2 \mathrm{GeV}) &=& -2.26(34) \times 10^{-3} \mathrm{GeV}^4 \nonumber \\
m_l\langle \overline{\psi}\psi \rangle_l^{\overline{MS}}(2 \mathrm{GeV}) &=& -7.63(4) \times 10^{-5} \mathrm{GeV}^4 .
\label{eq:mppval}
\end{eqnarray}
The ratio of the two values above is slightly more accurate 
than a naive combination, giving 29.6(4.3).  

Using the precise determinations for 
light quark masses now available from lattice QCD 
we can finally obtain condensate values. 
We take $m_s^{\overline{MS}}(2 \,\mathrm{GeV})$ = 92.2(1.3) 
MeV~\cite{Davies:2009ih, McNeile:2010ji}
and $m_s/m_l = 27.41(23)$~\cite{Bazavov:2009tw, milcreview}. 
These give:
\begin{eqnarray}
\langle \overline{s}s \rangle^{\overline{MS}}(2 \,\mathrm{GeV}) &=& -0.0245(37)(3) \,\mathrm{GeV}^3 \nonumber \\ 
&=& -(290(15) \,\mathrm{MeV})^3 \nonumber \\
\langle \overline{l}l \rangle^{\overline{MS}}(2 \,\mathrm{GeV}) &=& -0.0227(1)(4) \,\mathrm{GeV}^3 \nonumber \\ 
&=& -(283(2) \,\mathrm{MeV})^3,
\label{eq:ppval}
\end{eqnarray}
where the second error for each condensate in $\mathrm{GeV}^3$ comes 
from the error in the quark masses. 

For the ratio of strange to light condensate we have: 
\begin{equation}
\frac{\langle \overline{s}{s} \rangle^{\overline{MS}}(2 \,\mathrm{GeV})}{\langle \overline{l}l \rangle^{\overline{MS}}(2 \,\mathrm{GeV})} = 1.08(16)(1) ,
\label{eq:ppratval}
\end{equation}
where the first error comes from $R_s/R_l$ and has the 
error budget given in Table~\ref{tab:errbudget} and the 
second error comes from the strange to light quark mass ratio. 

\subsubsection{Approach of $R$ to the chiral limit}
\label{subsub:rchiral}
The relationship of the light quark condensate 
to the chiral condensate is also important. 
$R_q$ is defined to have the value 
1 from the GMOR relation in the chiral limit but 
the results of Eq.~\ref{eq:rval} indicate that it approaches 
this limit from below as the light quark 
mass is reduced. Supporting evidence for this 
is found by studying the quantity $R_{\delta}$ derived 
from the combination of condensates used by the 
HOTQCD collaboration in their study of finite 
temperature QCD~\cite{Bazavov:2011nk}. We define 
$R_{\delta}$ by: 
\begin{equation}
R_{\delta} = \frac{4m_l}{f_{\pi}^2M_{\pi}^2} \frac{\langle\overline{\psi}\psi_l\rangle - \frac{m_l}{m_s} \langle \overline{\psi}\psi_s \rangle}{1-\frac{m_l}{m_s}} . 
\label{eq:rdelta}
\end{equation}
The quadratic divergence with lattice spacing cancels between 
the two condensates because it is linear in the quark 
mass to all orders in perturbation theory. The non-divergent perturbative 
contributions proportional to the cube of the quark 
mass are completely negligible here, from the perturbative 
analysis in section~\ref{subsec:pt}, and so we do 
not need to include them in making $R_{\delta}$ a 
physical quantity. $R_{\delta}$ can then simply  
be calculated from the raw data in Table~\ref{tab:valence}.   

\begin{center}
\begin{figure}[ht]
\includegraphics[width=0.9\hsize]{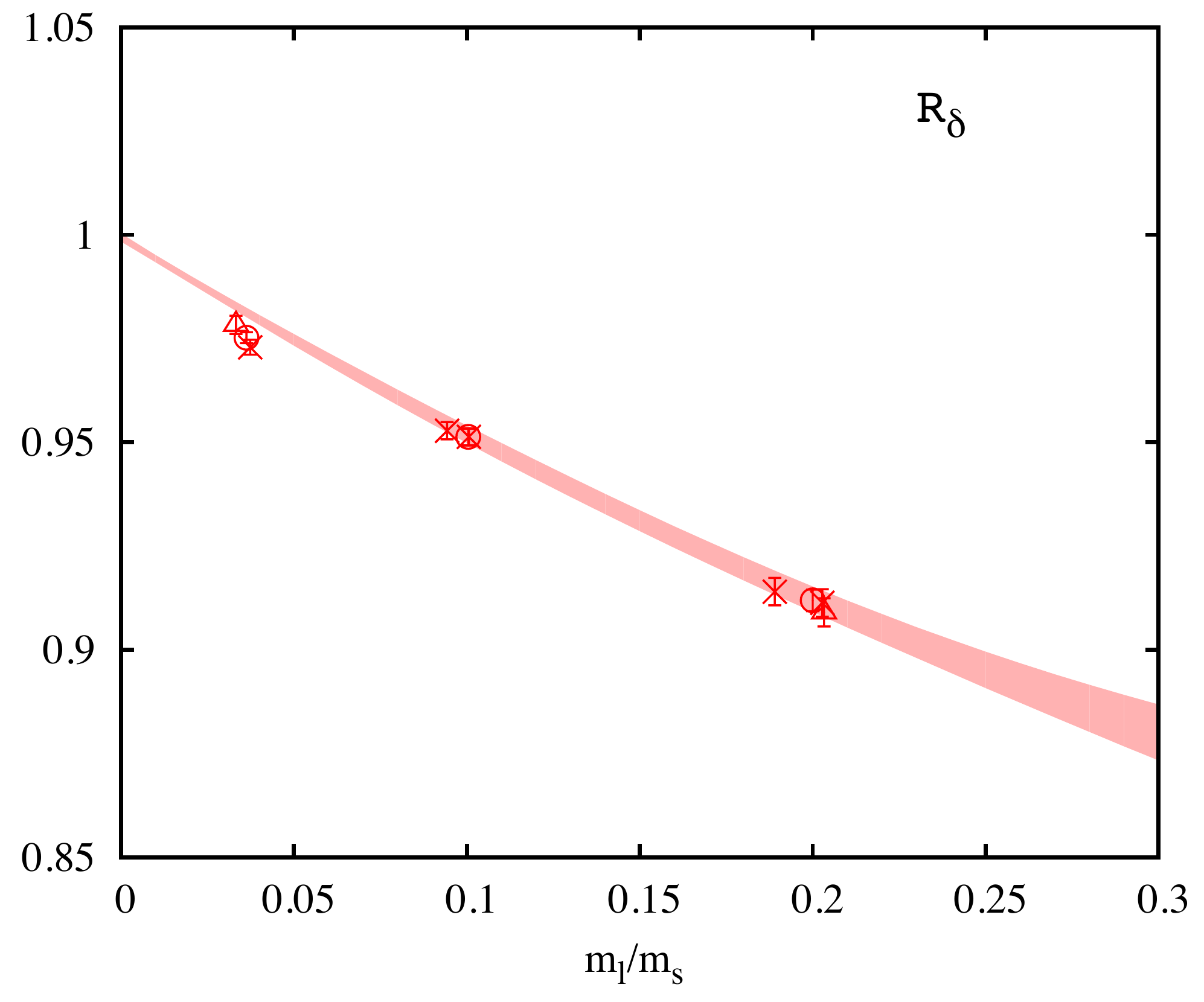}
\caption{$R_{\delta}$ as a function of 
$m_l/m_s$ at three values of the lattice 
spacing. Points show the raw lattice 
results: the crosses are from very coarse 
lattices (sets 1 and 2 with two 
values of $m_s$ on each and set 3), open circles from 
coarse lattices (sets 4, 5 and 6) and
open triangles from fine sets 7 and 8. 
The shaded band gives the results of a simple 
fit incorporating discretisation and 
finite volume effects as described in the text. 
}
\label{fig:rdelta}
\end{figure}
\end{center}

A plot of $R_{\delta}$ against $m_l/m_s$ is shown 
in Fig.~\ref{fig:rdelta}. In the $m_l \rightarrow 0$ 
limit on an infinite volume $R_{\delta} \rightarrow 1$ as $R_l$ does. 
$R_{\delta}$ can be determined more precisely
than $R_l$, however, 
because of the nonperturbative cancellation 
of the power divergence and it
clearly approaches 1 from below. $R_{\delta}$  
differs from $R_l$ by a term which is proportional 
to $m_l/m_s$ and to the difference between 
$\langle \overline{s}s\rangle$ and $\langle \overline{l}l \rangle$. 
Both the dependence of $R_l$ on $m_l$ and 
the difference between $R_{\delta}$ and $R_l$ 
then contribute to the slope with $m_l$ seen in Fig.~\ref{fig:rdelta}.
We cannot separate them and therefore unambiguously identify 
the slope of $R_l$ with $m_l$. We can however use this 
for a consistency check. 
 
We fit $R_{\delta}$ to the simple form:
\begin{eqnarray}
R_{\delta} &=& 1.000(1) + c_1(1+c_2a^2+c_3a^4)\frac{m_l}{m_s} \nonumber \\
&+& c_4\left(\frac{m_l}{m_s}\right)^2 + c_5e^{-m_{\pi}L}. 
\label{eq:rdeltafit}
\end{eqnarray}
This allows for linear and quadratic terms in $m_l$ with 
discretisation errors. We take priors on $c_1$, $c_3$ and 
$c_4$ to be 0.0(1.0) and prior on $c_2$ to be 0.0(0.3) 
consistent with $\alpha_sa^2$ behaviours. The final term 
allows for finite volume effects dependent on the combination 
$m_{\pi}L$. As for our fit to $R_l$, 
we take the prior 
on $c_5$ to be 0.0(0.2) for consistency with Fig.~\ref{fig:finitevol}. 

The fit gives $\chi^2/\mathrm{dof}$ of 0.75 for 10 degrees of 
freedom and a physical slope, $c_1$, of -0.51(4). This value 
is consistent with 
the difference between $R_l$ and 1 in Eq.~\ref{eq:rval}, 
and indeed with the difference between $R_s$ and 1. 
This consistency between the results from $R_{\delta}$ and 
$R_l$ indicates that the difference between $\langle \overline{s}s\rangle$ 
and $\langle \overline{l}l \rangle$ (which would upset 
this consistency) cannot be large. 
This is indeed what we also find in Eq.~\ref{eq:ppratval}. 

\subsubsection{The chiral susceptibility}
\label{subsub:susc}  
A further quantity that 
is of particular interest in studies of QCD at finite 
temperature is the chiral susceptibility for a quark 
of flavor $f$:
\begin{equation}
\chi_f = \frac{\partial}{\partial m_f} (-\langle {\overline{\psi}}\psi_f \rangle ). 
\label{eq:susdef}
\end{equation}
We give results here for the chiral susceptibility for zero temperature QCD 
to fill out the 
physical picture of the condensate. 
From differentiation of the path integral for the condensate 
it is clear that the chiral susceptibility is given by 
the flavor-singlet scalar correlator.  
It is convenient to split this into two contributions 
which we call $\chi_q$ and $\chi_g$ 
\footnote{In~\cite{Bazavov:2011nk} 
these are called $\chi_{conn}$ and $\chi_{disc}$.}.
$\chi_q$ comes from two scalar operators connected by quark lines, 
which is the flavor-nonsinglet scalar meson correlator. 
$\chi_g$ comes from 
two scalar operators connected 
only by gluons, in which the disconnected contribution 
is cancelled. 
\begin{eqnarray}
\chi &=& \chi_{q} + \chi_{g}  \\
\chi_{q} &=& \frac{1}{N_t} \sum_n \Tr \left[ M^{-1}_{0n} M^{-1}_{n0} \right]                    \nonumber \\
\chi_{g} &=& -\frac{1}{N_t^2V} \left(\langle (\Tr M^{-1})^2 \rangle - \langle \Tr M^{-1} \rangle^2 \right)\nonumber .
\label{eq:susdef2}
\end{eqnarray}
The factors of number of tastes, $N_t$, above are specific 
to naive/staggered quarks. 

\begin{table}
\caption{
Contributions to the chiral susceptibility, defined in 
Eq.~(\ref{eq:susdef2}) for very coarse set 1 and coarse 
set 4 for light ($m_l = m_s/5$) and strange quark masses. 
}
\label{tab:sus}
\begin{ruledtabular}
\begin{tabular}{llll}
Set & $ma$ & $a^2 \chi_q$ & $a^2\chi_g$ \\
\hline
1 & 0.013 & 0.54296(36) & 0.045(14) \\
  & 0.0688 & 0.45359(6) & 0.021(7) \\
\hline
3 & 0.01044 & 0.50850(18) & 0.032(10) \\
  & 0.0522 & 0.46231(3) & 0.014(4) \\
\hline
\end{tabular}
\end{ruledtabular}
\end{table}

\begin{center}
\begin{figure}[ht]
\includegraphics[width=0.9\hsize]{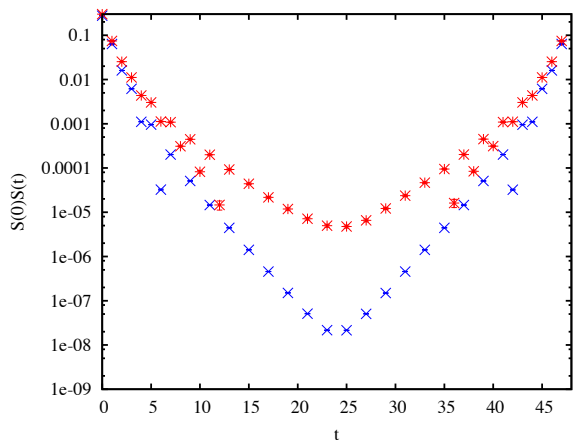}
\caption{
Results for the quark-line connected scalar correlator 
on set 1: $s$ quarks (blue crosses) and light 
quarks with $m_l=m_s/5$ (red bursts). The sum over time of
this correlator is $\chi_q$. 
Negative values of the correlator are not plotted 
on this log scale. 
}
\label{fig:scalar-conn}
\end{figure}
\end{center}

\begin{center}
\begin{figure}[ht]
\includegraphics[width=0.9\hsize]{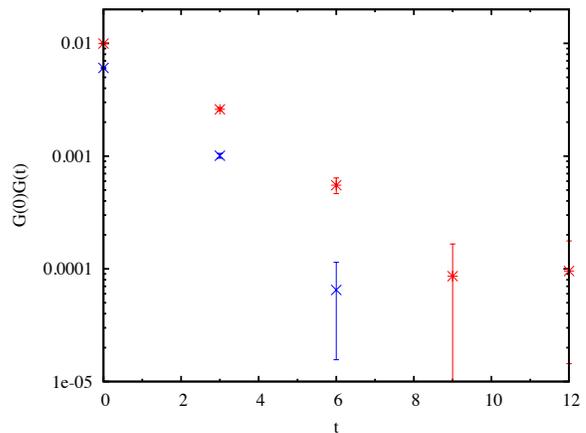}
\caption{
Results for the gluon-connected contribution 
to the scalar correlator 
on set 1 for $s$ quarks (blue crosses) and light 
quarks with $m_l=m_s/5$ (red bursts). The sum over time of
this correlator is $\chi_g$. 
}
\label{fig:glue}
\end{figure}
\end{center}

$\chi_{q}$ is readily calculated by generating quark 
propagators with the same random wall source of noise as that 
used for the $\pi$ and $\eta_s$ mesons, but 
patterned with phases that are -1 on all odd sites on 
an even-odd partitioning of the lattice. 
We then combine one of these propagators with the matching 
one used in the $\pi/\eta_s$ meson, again multiplying 
the odd sink sites with a phase of -1. 
Summing over spatial sites at each time slice gives the 
flavor-nonsinglet scalar correlator. Examples 
are shown in Fig.~\ref{fig:scalar-conn}. 
Summing this correlator over time slices gives $\chi_q$. 
Results for $\chi_q$ for light and strange quarks on 
sets 1 and 4 are given in Table~\ref{tab:sus}. 

$\chi_g$ can be estimated from our existing results for 
the light and strange condensates. Because we have 16 
time sources for our propagators we can determine 
the correlation between $\Tr M^{-1}$ operator that 
are $n$ time slices apart where $n$ is a multiple of 3 for set 1 
and a multiple of 4 for set 4. This gives a correlation 
function, for example that shown in Figure~\ref{fig:glue}. 
$\chi_g$ is then the sum over time slices of this 
correlation function. We can estimate $\chi_g$ in several 
ways. Our central result comes from estimating an effective 
mass from the early time slices that dominate $\chi_g$ 
and where we have a strong signal. We can then reconstruct 
an estimated correlation function and sum over it.  
We can also simply sum over the correlator for the time 
slices that we have and 
multiply by 3 or 4 as appropriate. From this range of 
methods we estimate the error in $\chi_g$ as 30\%.  
Values are given in Table~\ref{tab:sus}. $\chi_g$ 
is much smaller than $\chi_q$. 

\begin{center}
\begin{figure}[ht]
\includegraphics[width=0.9\hsize]{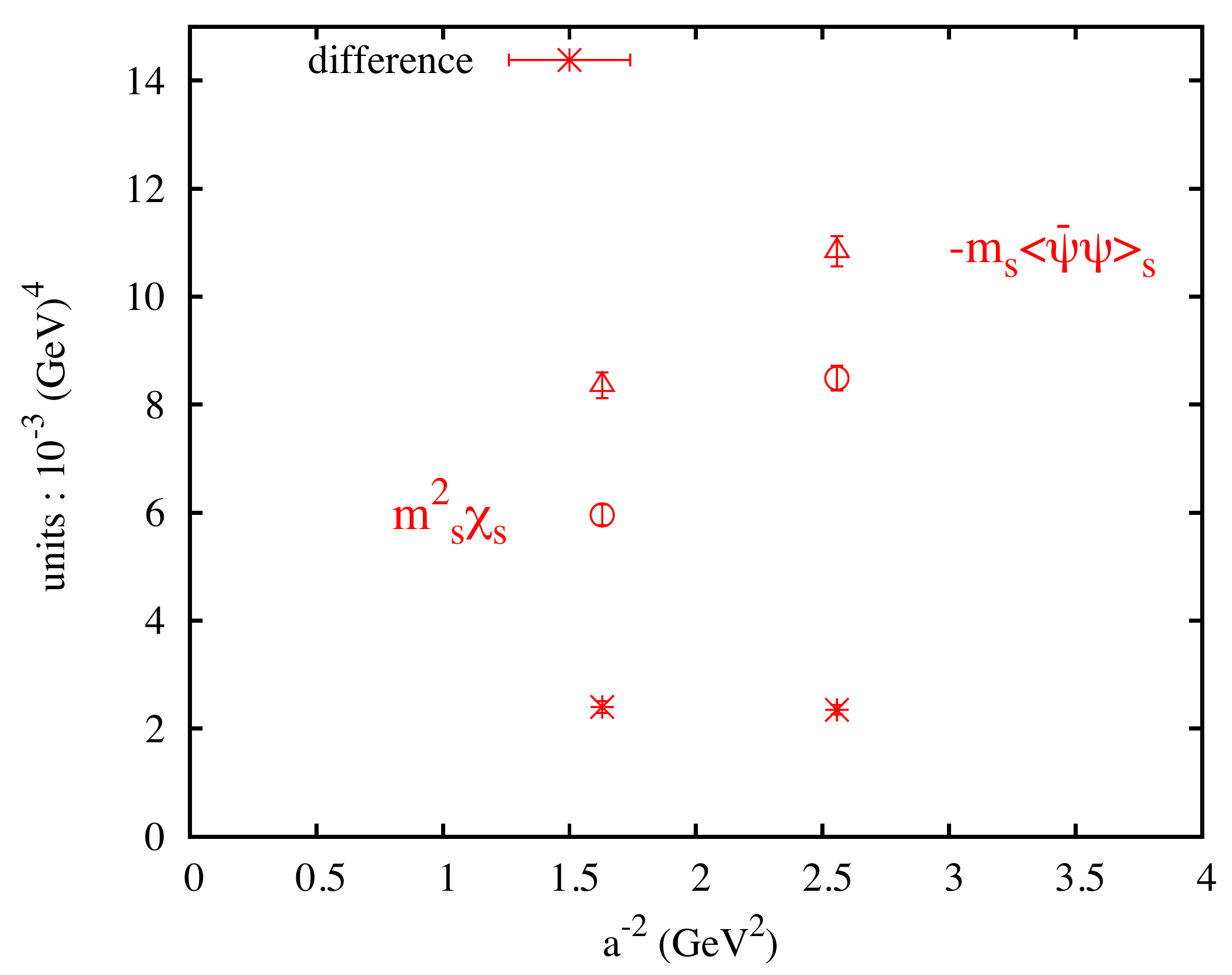}
\includegraphics[width=0.9\hsize]{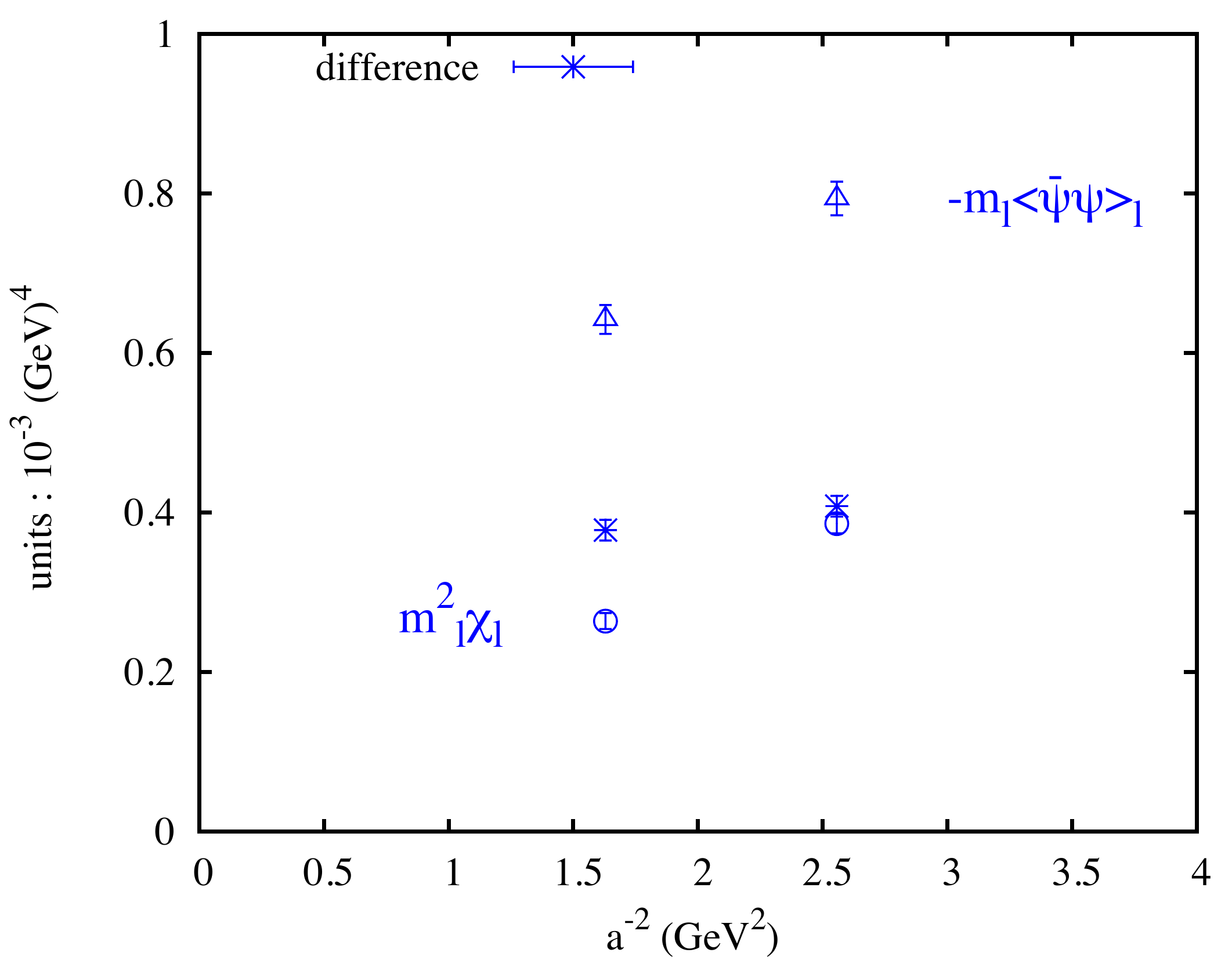}
\caption{
Results for $m_f^2\chi_f$ (open circles) compared to 
$m_f\langle \overline{\psi}\psi_f \rangle$ (open triangles) 
as a function of the square of the inverse 
lattice spacing, for $s$ quarks (top) and 
light quarks with $m_l=m_s/5$ (lower plot).  
Results are for sets 1 and 4. 
Bursts show the difference of these 
two quantities. 
}
\label{fig:sus}
\end{figure}
\end{center}

The quantities that are almost Renormlisation-Group 
invariant, that we can compare, are $m_f^2\chi_f$
and $m_f(-\langle \overline{\psi}\psi_f \rangle$. 
This is done in Figure~\ref{fig:sus}. 
Both of these quantities contain the same power 
divergence (proportional to $m_f^2a^{-2}$) 
in the lattice spacing. In the susceptibility 
this divergence largely comes from $\chi_q$. 
The nondivergent perturbative 
contributions to the two quantities will 
be different but, as we have seen, 
they are very small and we ignore them here. 

We find that the difference between $m_f^2\chi_f$ and 
$-m_f \langle \overline{\psi}\psi_f \rangle$, also 
plotted in Figure~\ref{fig:sus}, does not 
depend on the lattice spacing within errors.  
We simply average over the results at the two 
values of the lattice spacing to obtain physical 
results for
\begin{equation}
m_f \left( 1 - m_f \frac{\partial}{\partial m_f} \right) \langle -\overline{\psi}\psi_f \rangle.  
\label{eq:xdef}
\end{equation}
The values are: $2.18(7) \times 10^{-3} \mathrm{GeV}^4$ for $s$ quarks 
and $3.93(9) \times 10^{-4} \mathrm{GeV}^4$ for light 
quarks with $m_l = m_s/5$. 
Comparison of these numbers with the physical results for 
$m_f\langle \overline{\psi}\psi_f \rangle$ given in 
Eq.~(\ref{eq:mppval}) allows us to determine the value and sign of 
$m_f^2\chi_f$. For $s$ quarks the comparison is straightforward 
and we find:
\begin{equation}
m_s^2\chi_s = 0.08(35) \times 10^{-3}\mathrm{GeV}^4, 
\label{eq:mchis}
\end{equation}
consistent with zero. 

To obtain a value 
for $m_l^2\chi_l$ 
at $m_l=m_s/5$  we need a result
for $m_l\langle \overline{\psi}\psi_l \rangle$ 
at this mass. Our fit in Sec.~\ref{subsub:fits} gives a result for 
$R_{l=s/5}$ of 0.915(26). At this value of 
$m_l$ we have $M_{\pi}$ = 315 MeV and can estimate 
$f_{\pi}$ = 145 MeV from our results in Table~\ref{tab:valence}. 
Then $m_{l=s/5}\langle \overline{\psi} \psi_{l=s/5} \rangle = 4.78(20) \times 10^{-4} \mathrm{GeV}^4$. 
In the error we have included 
an interpolation error for each of the mass and decay constant of 1\%. 
Then, subtracting the result for the difference of Eq.~\ref{eq:xdef} 
at $m_l=m_s/5$ given above, we have 
\begin{equation}
m_{l=s/5}^2\chi_{l=s/5} = 0.85(22) \times 10^{-4}\mathrm{GeV}^4, 
\label{eq:mchil}
\end{equation}
which is a small positive slope. 

Our results are then consistent with a slope in 
$\langle \overline{\psi} \psi_f \rangle$ with $m_f$ that is positive 
at $m_s/5$ but may decrease or even become negative by $m_s$. 
If the slope at $m_s/5$ remained constant for larger $m_f$ 
it would give a total change in $\langle \overline{\psi}\psi_f \rangle$ 
of $3\times 10^{-3}$ between $m_s/5$ and $m_s$ which is 
not inconsistent with the change of $1.8(3.8) \times 10^{-3}$ 
that we find in Eq.~\ref{eq:ppval}. 

\section{Discussion}
\label{sec:discussion}

We have determined a physical value for the strange quark 
condensate from lattice QCD for the first time. 
This required both nonperturbative lattice QCD results 
and a perturbative determination of the power 
divergent contribution through $\mathcal{O}(\alpha_s)$. 
The calculation relies on the good chiral properties 
of staggered quarks to control the form of the 
power divergence and the numerical speed and 
small discretisation of the Highly Improved Staggered 
quark formalism allow very precise results to be obtained 
at several values of the lattice with light $u/d$ sea 
quark masses. We proceed by tracking deviations from 
the GMOR relation, making extensive use of 
our previous work determining the physical properties 
of the $\eta_s$ meson, to obtain the strange quark 
condensate.  
Our best result comes from gluon field 
configurations that include $u$, $d$, $s$ 
and $c$ quarks in the sea. The condensate 
is given in the $\overline{MS}$ scheme at a 
scale of 2 GeV. The evolution equation required 
to run $m\langle \overline{\psi}\psi\rangle$
to other scales, since it is not RG-invariant, 
is given in Appendix~\ref{appendix:appb}. 
We obtain a very consistent result for the 
strange quark condensate from 
independent calculations that include 
2+1 favors of sea quarks, as discussed in 
Appendix~\ref{app:lattice2}. 

Our value is $-(290(15) \, \mathrm{MeV})^3$ giving 
a ratio of strange to light condensates of 1.08(16). 
Earlier results come from sum rules of various kinds. 
These show significant variation and often have no 
estimate of the error associated with the value. 
Narison~\cite{Narison:2002hk} gives a compilation 
with a final value for the ratio of strange to 
light condensates in the $\overline{MS}$ scheme at a 
scale of 2 GeV of 0.75(12). A value of 0.74(3) is 
quoted from baryon mass splittings in~\cite{Albuquerque:2009pr}.  
Finite energy sum rules in the kaon sector 
give a ratio 0.6(1) in~\cite{Dominguez:2007my}. 
More recently Maltman~\cite{Maltman:2008na} uses 
sum rules for the ratio of decay constants $f_{B_s}/f_B$ 
along with the 2007 lattice QCD average for this ratio of 
1.21(4)~\cite{Gray:2005ad, Bernard:2007zz} 
to obtain $\langle s\overline{s} \rangle/\langle l\overline{l}\rangle_{\overline{MS}} (2 \mathrm{GeV})$ = 1.2(3). This updates an 
earlier result of 0.8(3) from Jamin~\cite{Jamin:2002ev} 
which used a quenched lattice 
QCD result for $f_{B_s}/f_B$ of 1.16(4). 
The current lattice QCD world 
average for $f_{B_s}/f_B$ is 1.20(2)~\cite{Davieslat11}.

\begin{center}
\begin{figure}[ht]
\includegraphics[width=0.8\hsize]{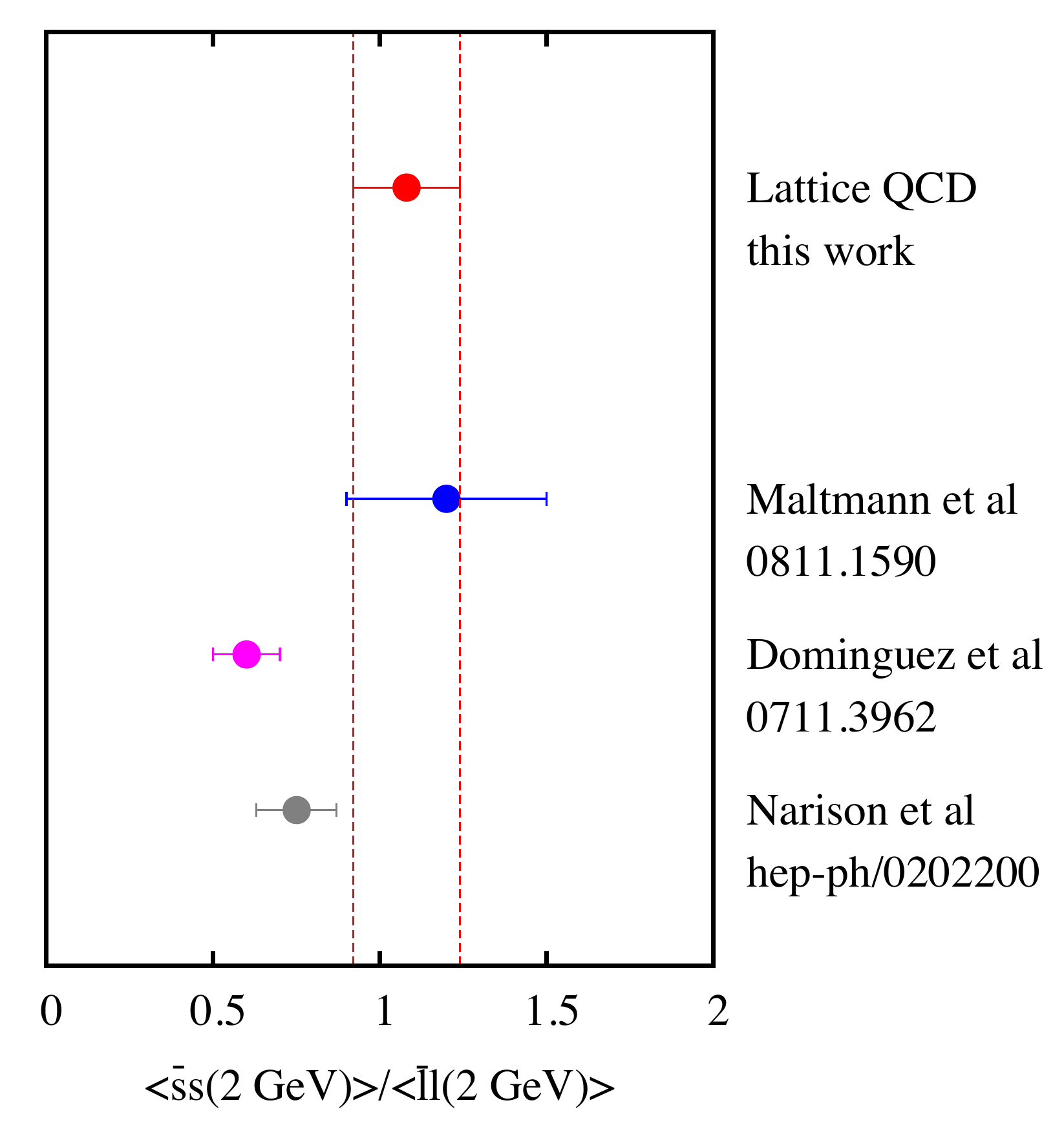}
\caption{
A comparison of results for the ratio of 
strange to light condensates in the $\overline{MS}$ 
scheme at 2 GeV. }
\label{fig:condcomp}
\end{figure}
\end{center}

Fig.~\ref{fig:condcomp} compares our result 
for the strange to light condensate ratio with the 
results from sum rules discussed above. 
Our central value lies between the sum rules results, 
being in agreement with the larger value of~\cite{Maltman:2008na} 
but only in marginal agreement with the lower 
values of~\cite{Dominguez:2007my}. A value below 
0.6 is ruled out by our results at the $3\sigma$ level. 
Our value is 
more accurate than the result 
derived from $f_{B_s}/f_B$ and has the advantage 
over all the sum rules results that it is a direct 
determination from QCD and has a full error budget 
(Table~\ref{tab:errbudget}).   

We obtain a very accurate value for the light 
quark condensate,  
giving $-(283(2)\,\mathrm{MeV})^3$. 
We can distinguish the ratio $R_l$ at the 
physical light quark mass from that of 
1 in the chiral limit 
using our results in Eq.(~\ref{eq:rval}).
Defining $\delta_{\pi}$ from~\cite{Jamin:2002ev} 
\begin{equation}
R_l = 1-\delta_{\pi}
\end{equation}
we obtain a value $\delta_{\pi}=0.015(5)$. 
This is somewhat lower than the value of 0.047(17) 
estimated in~\cite{Jamin:2002ev}, although in 
agreement within $2\sigma$. It is not in agreement with 
the somewhat larger number of 0.06(1) obtained in~\cite{Bordes:2010wy}. 
Our result implies 
a value for the combination $(2L^r_8-H^r_2)$ of low energy 
constants from the chiral Lagrangian that is a factor of 
three lower than that used in~\cite{Jamin:2002ev}. 
The value is 3$\sigma$ larger than zero, however. 

Note that we do not expect the value of the 
light quark condensate to agree with that of the 
chiral (zero quark mass) condensate, $\Sigma$.  
The relationship between them is: 
\begin{equation}
\Sigma(2\,\mathrm{GeV}) = \frac{-\langle \overline{l} l \rangle(2\,\mathrm{GeV})}{R_{l,phys}}\frac{f^2}{f_{\pi}^2}\frac{(M^2/m)_{m=0}}{M_{\pi}^2/m_{l,phys}}
\label{eq:sigmacalc}
\end{equation}
Here $f$ is the decay constant in the zero quark mass limit and 
$M^2/m$ is the ratio of the square of the pion mass to the light 
quark mass in the same limit. 
$f_{\pi}/f$ can be determined from chiral extrapolation of lattice 
QCD results. 
For example, a recent accurate calculation~\cite{Borsanyi:2012zv} 
gives
$f_{\pi}/f=1.0627(28)$ including 2+1 flavors of sea quarks. 
A preliminary analysis based on the results 
given here for 2+1+1 sea flavors in Table~\ref{tab:valence} gives 1.056(1), 
in acceptable agreement.  
From the figures in~\cite{Borsanyi:2012zv} we estimate 
$(M^2/m)_{m=0}/(M_{\pi}^2/m_{l,phys})$ as 1.02. 
Combining these factors, along with $R_l$, into Eq.~\ref{eq:sigmacalc}
makes clear that we expect a 3\% difference between the magnitudes of  
$\Sigma^{1/3}$ and $(\langle \overline{l} l \rangle)^{1/3}$, 
dominated by the effect of $f_{\pi}/f$ (so that $\Sigma$ is smaller). 
This is entirely consistent, assuming no difference between 
2+1 and 2+1+1 flavors of sea quarks, with the fact that we obtain 
$\langle \overline{l} l \rangle (2\mathrm{GeV}) = -(283(2)\,\mathrm{MeV})^3$ 
and~\cite{Borsanyi:2012zv} obtain $\Sigma(2\mathrm{GeV}) = (272(2)\,\mathrm{MeV})^3$ 
from a chiral analysis. 

Other methods for determining $\Sigma$ are not as accurate, but 
in reasonable agreement. 
We quote here two recent examples. 
JLQCD/TWQCD give a result of $(234(17)\,\mathrm{MeV})^3$~\cite{Fukaya:2010na} from 
the eigenvalue spectrum of overlap quarks with $u$, $d$ and $s$ 
quarks in the sea. The ETM collaboration 
give $(299(38)\,\mathrm{MeV})^3$~\cite{Burger:2012ti}    
from fits to the Landau gauge quark propagator with 
$u$ and $d$ quarks in the sea. Additional lattice results 
for $\Sigma$ are collected in~\cite{flag}.  

Our analysis has implications for other calculations. For example: 
\begin{itemize}
\item {Finite temperature determinations of the chiral 
phase transition in QCD use an order parameter 
based on the light quark condensate. 
A nonperturbative subtraction is made with the 
aim of removing the power divergent pieces proportional 
to $ma$ and 
with the assumption that the higher order ($(ma)^3$) 
terms are negligible. For example the HOTQCD
collaboration uses an order parameter~\cite{Bazavov:2011nk} 
for visualising the transition (fits to find 
the transition temperature also include other results) which 
is the ratio between non-zero and zero temperature of  
\begin{equation}
\langle \overline{\psi}\psi \rangle_l - \frac{m_l}{m_s} \langle \overline{\psi}\psi \rangle_s .
\label{eq:hotqcd-order}
\end{equation}
This quantity becomes 
\begin{equation}
\langle \overline{\psi}\psi \rangle_{l,NP} - \frac{m_l}{m_s} \langle \overline{\psi}\psi \rangle_{s,NP} 
\label{eq:hotqcd-order2}
\end{equation}
if we assume only the presence of a power divergence linear 
in $ma$. Our analysis shows that this is a good assumption. 
For example, the difference between subtracting only 
terms linear in $ma$ at tree level and including terms 
cubic in $ma$ is 0.2\% for the strange condensate on the coarsest 
HOTQCD lattices (i.e. those with largest $m_sa$ values). 
An alternative might be to calculate 
$(1-m\partial/\partial m)\langle \overline{\psi}\psi \rangle$ 
as discussed in Sec.~\ref{subsub:susc}. The quark-line 
connected piece of this can be calculated directly by 
combining the expression for $\chi_q$ in Eq.~\ref{eq:susdef2} 
and the expression for $\langle \overline{\psi}\psi \rangle$ 
in Eq.~\ref{eq:condid}. The combination becomes~\cite{Kilcup:1986dg}: 
\begin{equation}
\left( 1 - \frac{\partial}{\partial m} \right)_{\mathrm{conn}} \langle -\overline{\psi} \psi \rangle = 2m \sum_{n \,\mathrm{even}} \Tr |M^{-1}_{0n}|^2.
\label{eq:comb}
\end{equation}  
This can clearly be generalised to a sum over even source sites, 
implemented with a partial random wall. 
When combined with the quark-line disconnected piece $\chi_g$ 
this gives a physical quantity without power divergent pieces which 
is close to the value of the condensate itself.  
}
\item{The comparison of heavy-light current-current correlators 
to continuum QCD perturbation theory can be used to normalise 
heavy-light currents in lattice QCD. The light quark condensate 
appears in this comparison and the results given here will 
enable us to improve the analysis in~\cite{Koponen:2010jy}. This is underway. 
}
\item{ 
Some recent papers~\cite{Brodsky:2010xf} have speculated 
that the quark condensate may only be non-zero inside hadrons. 
A much smaller value outside hadrons would significantly ameliorate
the fine tuning problem associated with the cosmological 
constant. 
This suggestion appears to be in conflict with 
direct calculations of quark condensates as vacuum 
expectation values described here. }
\end{itemize}

\section{Conclusions}
\label{sec:conclusions}
We give the first direct determination of the 
strange quark condensate from lattice QCD, 
having demonstrated how to extract a well-defined 
physical value from lattice results that contain 
a power divergence as the lattice spacing goes 
to zero. Our results include a calculation through 
$\mathcal{O}(\alpha_s)$ in lattice QCD perturbation 
theory of the perturbative contribution to the 
condensate, part of which is the power divergence. 
The calculation relies on the good chiral properties 
of staggered quarks to control the form of the 
power divergence and the numerical speed and 
small discretisation of the Highly Improved Staggered 
quark formalism to obtain precise results at 
multiple lattice spacings and light quark masses. 
Our results include values at physical light 
quark masses. 

We obtain a value for the strange quark 
condensate in the $\overline{MS}$ scheme 
at 2 GeV of $-(290(15)\,\mathrm{MeV})^3$. We give a full error budget 
for this result in Table~\ref{tab:errbudget}, 
the main sources of error being those 
associated with fitting and subtracting the 
remaining power divergence. 
The result includes $u$, $d$, $s$ and $c$ quarks 
in the sea but we get good agreement with this value from 
independent calculations that include $u$, $d$ and 
$s$ sea quarks only.  

The value we obtain for the corresponding 
light quark condensate (where $m_l = (m_u+m_d)/2$) 
is $-(283(2)\,\mathrm{MeV})^3$. Note that is 
significantly different from the value for the condensate 
in the chiral limit. The ratio of our light quark 
condensate to a recent lattice QCD value for the chiral condensate 
from~\cite{Borsanyi:2012zv} is 1.13(3), consistent with 
the behaviour of meson masses and decay constants approaching 
the chiral limit.  

We have shown that the ratio of four times the quark mass times 
condensate divided by the square of the meson mass times
decay constant approaches the GMOR value (of 1) 
from below as $m_l \rightarrow 0$. At the physical 
light quark mass the value is 1.5\% below 1, and at 
the strange mass it is 57\% of 1. 

Our result for the ratio of the strange condensate 
to the light quark condensate is 1.08(16). 
This sits in the middle of the spread of results 
from QCD sum rules but provides significant additional 
information because it is a direct determination with a 
full error budget. 
The result will have impact on a number of other 
calculations both in the continuum and in lattice QCD. 
Some of the numerical techniques used here will 
be useful for determinations of, for example, the 
strangeness content of the pion or nucleon. 

{\bf Acknowledgements} We are grateful to Eduardo Follana, Elvira Gamiz, 
Matthias Jamin, Jack Laiho, Doug Toussaint 
and Roman Zwicky for useful discussions and to the MILC collaboration 
for the use of their 2+1+1 gauge field configurations. The 
calculations described here were performed on the Darwin 
supercomputer of the Cambridge High Performance Computing 
service as part of the DiRAC facility 
jointly funded by STFC, BIS and the Universities of Cambridge 
and Glasgow. We are grateful to the Darwin support staff for assistance. 
Our work is supported by the Royal Society, the Wolfson 
Foundation, the Science and Technology Facilities Council, 
the National Science Foundation and 
the U.S. Department of Energy (under Contract DE-AC02-98CH10886). 

\appendix

\section{Condensates from correlators}
\label{appendix:appa}
Eq.~(\ref{eq:ppbardef}) relates the zero-momentum pseudoscalar propagator to the scalar quark-condensate and is a well-known relationship~\cite{Kilcup:1986dg}. 
The relationship is true (for the HISQ formalism) even on 
a single gluon configuration. 
Here, for completeness, we give a simple derivation, using the equivalent 
naive quark formalism. 

The contribution to the propagator from a single gluon configuration is given by 
\begin{equation}
   G_\mathrm{ps} \equiv \sum_x\mathrm{Tr}\left[\gamma_5 
   \left(\frac{1}{D\cdot\gamma + m}\right)_{0x} \gamma_5
   \left(\frac{1}{D\cdot\gamma + m}\right)_{x0} \right],
\end{equation}
where $D$ is the gauge-covariant derivative, and the trace is over spin and color indices. The contribution to the scalar quark-condensate is given by
\begin{equation}
   S \equiv -\mathrm{Tr}\left(\frac{1}{D\cdot\gamma +
    m}\right)_{00}.
\end{equation}
To extract the relationship, we multiply by the unit matrix under the trace in the condensate:
\begin{align}
S &=   -\sum_{xy}\mathrm{Tr}\left[
\left({-D\cdot\gamma + m}\right)_{0x}
\left(\frac{1}{-D\cdot\gamma + m}\right)_{xy} \times \right. \notag \\
&\quad\quad\quad\quad\quad
\left.\left(\frac{1}{D\cdot\gamma + m}\right)_{y0}\right] \notag \\
 &= -\sum_{x}\mathrm{Tr}\left[
 \left({-D\cdot\gamma + m}\right)_{0x}
 \left(\frac{1}{-(D\cdot\gamma)^2 + m^2}\right)_{x0}\right]
\end{align}
Only the $m$~term in the numerator of the last expression survives the spinor trace since the other term results in traces of odd numbers of $\gamma$~matrices (which vanish). Consequently
\begin{align}\label{gmor-rel}
   S &= -m \mathrm{Tr}
    \left(\frac{1}{-(D\cdot\gamma)^2 + m^2}\right)_{00}
    \notag \\
    &= -m \sum_{x}\mathrm{Tr}\left[ 
    \left(\frac{1}{-D\cdot\gamma + m}\right)_{0x}
    \left(\frac{1}{D\cdot\gamma + m}\right)_{x0}\right]
    \notag \\
    &= -m \sum_{x}\mathrm{Tr}\left[ \gamma_5
    \left(\frac{1}{D\cdot\gamma + m}\right)_{0x}\gamma_5 
    \right. \times  \notag \\
    &\quad\quad\quad\quad\quad\quad\left.
    \left(\frac{1}{D\cdot\gamma + m}\right)_{x0}\right]
    \notag \\
    &= -m\,G_\mathrm{ps},
\end{align}
which is Eq.~(\ref{eq:ppbardef}). Since this relationship is true configuration-by-configuration, it must be true of the ensemble averages as well. 

Note that Eq.~(\ref{gmor-rel}) leads immediately to the GMOR relation (Eq.~(\ref{eq:gmor})).
To see this rewrite the pseudoscalar propagator 
in terms of its mesonic intermediate states. 
Only the pion contribution survives the $m\!\to\!0$ limit, 
since the effective decay constants for excited states 
all vanish in that limit (by the Ward identity). 
The pion contribution has an amplitude $a=f_{\pi}^2m_{\pi}^3/(4m)$ multiplied 
by an exponential decay in time, whose integral gives $1/m_{\pi}$. 

The analysis of the propagator above only works for 
quark actions that have a $\gamma_\mu$ piece and a 
scalar piece (and nothing else), and where those 
two pieces commute with each other. The commuting is essential if you want 
\begin{equation}
  ( -D\cdot\gamma + m)(D\cdot\gamma + m) 
  = -(D\cdot\gamma)^2 + m^2
\end{equation}
with only terms having an even number of $\gamma_\mu$s 
on the right hand side. So this proof does \emph{not} work 
for Wilson's lattice discretization of the quark action or 
similar formulations. 
On the other hand, it is true of staggered-quark formalisms such 
as HISQ.

Eq.(~\ref{eq:ppbardef}) also follows directly from the (integrated) 
axial Ward identity:
\begin{eqnarray}
&&\sum_x \langle (m_a + m_b) J^5_{ab}(x) (m_a+m_b) J^{5\dag}_{ab}(0) \rangle =  \hspace{15mm} \nonumber \\
&& \hspace{22mm} - \langle (m_a + m_b) (\overline{\psi_a}\psi_a + \overline{\psi_b}\psi_b) \rangle
\label{eq:intwi}
\end{eqnarray}
with $J^5_{ab} \equiv \overline{\psi_a}(x)\gamma_5 \psi_b(x)$.
This is exact on the lattice for lattice actions with sufficient 
chiral symmetry and again shows that is an identity, true configuration 
by configuration and for any $m_a$ and $m_b$. 
See~\cite{Frezzotti:2005gi} for a derivation using 
twisted mass quarks. 

Here we have used the cases $m_a=m_b$ and both equal to either 
$m_l$ or $m_s$ but we can derive 
from Eq.(~\ref{eq:intwi}) a relationship~\cite{Kilcup:1986dg} between  
correlators for the mixed Goldstone pseudoscalar made of light 
and strange quarks and the `diagonal' cases: 
\begin{eqnarray}
(am_l + am_s) \sum_t C_{K}(t) = &&(am_l)\sum_t C_{\pi}(t) \nonumber \\ 
+ &&(am_s)\sum_t C_{\eta_s}(t) . 
\label{eq:kpieta}
\end{eqnarray}
The left-hand side is then related to the sum of quark masses multiplied by
the sum of quark condensates. 
This does not add new information so we do not 
make use of this relationship except as a test of our correlators. 

\section{Condensates and the OPE}
\label{appendix:appb}
Condensates typically arise in the non-leading terms of operator-product expansions~(OPE). To illustrate, consider moments of two pseudoscalar densities composed of a heavy quark (mass~$M$) and a light quark (mass~$m\ll  M$) where the heavy quark fields are contracted with each other:
\begin{equation}
(m+M)^2 \int d\mathbf{x}\,dt\,t^n\,
J_5(\mathbf{x},t)J_5(0) \to \On
\end{equation}
where
\begin{equation}
   \On \equiv \int d\mathbf{x}\,dt\,t^n\,
   \psib(\mathbf{x},t)\gamma_5
   \,\frac{(m+M)^2}{D\cdot\gamma+M}\,\gamma_5
      \psi(0),
\end{equation}
and $\psi$ is the light-quark field. 
Lattice simulations of~$\vacexp{\On}$ can be used to 
determine the heavy quark's mass~\cite{Allison:2008xk}.
The  $(m+M)^2$~factor makes $\mathcal{O}^{(n)}$ independent of the ultraviolet regulator provided $n\!\ge\!4$; that is, lattice and continuum calculations should agree in the limit of zero lattice spacing.

Operator~$\On$ is also short-distance, dominated by length scales of order~$1/M$, provided the heavy-quark is sufficiently heavy and the light quarks have momenta small compared with~$M$. Consequently the~OPE implies that~$\On$ can be expressed in terms of a set of local operators in an effective theory, with cutoff scale~$\Lambda\!<\!M$,  and coefficient functions that {depend only upon physics between scales~$\Lambda$ and~$M$}:
\begin{align} \label{ope}
   \Mb^{n-4}\On &= \mathbf{1}^{(\Lambda)}\, 
   c(\Lambda/\Mb,\alphab_s,\mb/\Mb)
    \nonumber\\
   &+ \frac{(\mcond)^{(\Lambda)}}{\mb\Mb^3}\,
   d(\Lambda/\Mb,\alphab_s,\mb/\Mb) \\
   &+\cdots.\nonumber
\end{align}
where $\mathbf{1}^{(\Lambda)}$ is the unit operator and we have replaced~$\psib\psi$ by $m\psib\psi$, to simplify the coefficient function. Somewhat arbitrarily, we have chosen to express the right-hand side in terms of masses and couplings at scale $\mu\!=\!M(\mu)\!\equiv\!\Mb$:
\begin{equation}
   \Mb \equiv M(\Mb), \quad
   \mb \equiv m(\Mb), \quad
   \alphab_s \equiv \alpha_s(\Mb).
\end{equation}
The effective theory on the right-hand side of \eq{ope} could be, for example, lattice QCD with a lattice spacing $a\!=\!\pi/\Lambda$, or QCD with an~$\msb$ regulator and~$\mu\!=\!\Lambda$.

The coefficient functions~$c$ and~$d$ are perturbative when~$M$ is large, and analytic in~$\alphab_s$ and~$\mb/\Mb$\,\footnote{We are ignoring nonperturbative 
short-distance physics (for example, small instantons) which can contribute to 
coefficient functions but is typically nonleading.}. 
They can be computed using perturbative matching. For example, we can examine matrix elements of \eq{ope} between low-energy, on-shell light-quark states $\langle q|$ and $|q^\prime\rangle$. The unit operator drops out and \eq{ope} can be rearranged to give 
\begin{equation} \label{d-calc}
   d(\Lambda/\Mb,\alphab_s,\mb/\Mb) = \left(
   \frac{\mb\Mb^{n-1}\qexp{\On}}{\qexp{\mcond}^{(\Lambda)}}
   \right)_\mathrm{PQCD},
\end{equation}
where the right-hand side is computed order-by-order in perturbation theory. Since $\qexp{\cond}$ is independent of~$\Lambda$, $d$ is actually regulator independent: 
\begin{equation}
   d\!=\!d(\alphab_s,\mb/\Mb).
\end{equation} 
Knowing~$d$, one would then compute $c$ using perturbative expansions of the vacuum expectation values:
\begin{equation} \label{c-calc}\begin{split}
   c(&\Lambda/\Mb,\alphab_s,\mb/\Mb) = \\
   &\left(
   \frac{\vacexp{\On}}{\Mb^{4-n}} 
    - \frac{\vacexp{\mcond}^{(\Lambda)}}{\mb\Mb^3}\,
   d(\alphab_s,\mb/\Mb)\right)_\mathrm{PQCD},
      \end{split}
\end{equation}

\eq{c-calc} underscores the importance of avoiding normal-ordered operators in operator-product expansions. Each term on the right-hand side has infrared sensitive contributions that go like $m^3\log(m)$. These cancel between the two terms in \eq{c-calc}, order-by-order in perturbation theory; but this cancelation would have been ruined had we replaced $\psib\psi$ by the normal-ordered product~$:\psib\psi:$ in~\eq{ope} (and $c$ would no longer be perturbative).

It is also important to note that $\vacexp{\mcond}^{(\Lambda)}$ is \emph{not} cutoff independent, because of operator mixing with the unit operator~$m^4\mathbf{1}$, which implies that
\begin{equation} \label{cond-evol}
   \frac{d\vacexp{\mcond}^{(\Lambda)}}{d\log\Lambda}
   = \gamma_\mathrm{mix}(\alpha_s(\Lambda),m(\Lambda)/\Lambda)\,m^4(\Lambda).
\end{equation}
where
\begin{equation} \label{msb-evol2}
   \gamma_\mathrm{mix}(\alpha_s) = \frac{3}{2\pi^2} + O(\alpha_s).
\end{equation}
depends upon the regulator scheme beyond tree-level. This evolution is 
typically negligible for light quarks because of the $m^4$~factor.

\section{$\msb$ condensates from the lattice}
\label{appendix:appc}
The coefficient functions in operator-product expansions such 
as \eq{ope} are most conveniently computed using 
the $\msb$~regulator to define the operators on the 
right-hand side. On the other hand, the only technology 
available for determining the nonperturbative matrix elements 
needed in such analyses is lattice simulation, 
using the lattice ultraviolet regulator. 
To combine these techniques we must be able to 
convert lattice determinations of $\vacexp{\mcond}$, 
for example, into the equivalent $\msb$ matrix elements.

The relationship is again given by the operator product expansion:
\begin{align} \label{msb-cond} 
   (\mcond)^{(\mu)}_{\msb} &=  \mathbf{1}^{(a)}
    \frac{m^2}{a^2}f(\mu\leftrightarrow \pi/a) \notag \\
   &\quad+ (\mcond)^{(a)}_\mathrm{LQCD}\, h(\mu\leftrightarrow \pi/a) + \cdots
\end{align}
where the coefficient functions~$f$ and~$h$ can only depend upon physics between~$\mu$ and the lattice cutoff~$\pi/a$. In fact $h\!=\!1$ since the matrix elements in
\begin{equation} \label{msb-cond2}
      h = \frac{\qexp{\mcond}^{(\mu)}_{\msb}}{\qexp{\mcond}^{(a)}_\mathrm{LQCD}}
        = 1
\end{equation}
are~$\mu$ and~$a$ independent, and therefore $h$~must be a number (and~1 
is the correct number, from perturbation theory). The coefficient 
function~$f$ is computed order-by-order in perturbation theory 
using the expansions of the two condensates 
(computed with their respective regulators):
\begin{align} \label{msb-cond3}
   f
   &\equiv \frac{a^2}{m^2} \left(\vacexp{\mcond}^{(\mu)}_{\msb} - 
   \vacexp{\mcond}^{(a)}_\mathrm{LQCD}
   \right)_\mathrm{PQCD} \notag \\
   &= \sum_{n=0} f_n^{(0)}(a\mu)\, \alpha_{\msb}^n(\mu) \notag\\
   &\quad + (am)^2 \sum_{n=0} f_n^{(1)}(a\mu)\, \alpha^n_{\msb}(\mu)
\end{align}
The cancellation of all~$\log m$ terms between the two matrix elements in~$f$ is something of a miracle; it only works if the~$m$ in each case is precisely the~$m$ that multiplies $\bar\psi\psi$ in the action for that case
~\footnote{If this isn't the case, then coefficient function~$h$ will not equal one, but rather will be a series in~$\alpha_\msb$.}.
$\Delta_{PT}$ in eq.(~\ref{eq:deltapert}) is $(am)^2f$ calculated 
through $\mathcal{O}(\alpha_s)$. 

Additional terms appear in Eq.~\ref{msb-cond} from mixing with higher 
dimension condensates, such as the gluon condensate. 
These are suppressed by positive powers of $a$. 
For the gluon condensate the multiplier is $(ma)^2$ 
from chirality arguments. These terms then simply 
look like discretisation errors in $m\overline{\psi}\psi$ 
and are handed as part of the general treatment of 
those errors. 

\section{Lattice QCD calculation on $n_f=2+1$ gluon configurations}
\label{app:lattice2}
We
show here further results for the strange quark condensate 
from two contrasting calculations, both using HISQ quarks, that include 
$u$, $d$ and $s$ quarks in the sea, but no $c$ quarks. 
The first calculation uses sets of MILC configurations 
corresponding to 5 values of lattice spacing spanning a 
large range from 0.15 fm to 0.04 fm~\cite{Davies:2009tsa} 
and using the asqtad formalism for the sea quarks. The second
uses HOTQCD configurations~\cite{Bazavov:2011nk} and has more 
lattice spacing values (24 in total) but with only a limited 
number (9) having accompanying meson masses and 
decay constants. The sea quarks are included using the
HISQ formalism with $u/d$ sea quark masses close to 
the physical value. 
The second 
calculation corresponds to the zero temperature 
results 
generated to accompany a finite temperature analysis 
of the phase structure of QCD. This analysis needs 
many values of the lattice spacing for a fine-grained 
temperature scale, and the zero temperature results 
are needed to fix the QCD parameters. 
The quark condensate is an important 
order parameter at finite temperature but is also 
determined in~\cite{Bazavov:2011nk} on the zero 
temperature ensembles. 

\begin{table*}
\caption{
Raw (unsubtracted) values for the 
strange quark condensate along with $\eta_s$ masses 
and decay constants in lattice 
units calculated for valence masses given in column 4. 
The calculations use valence HISQ quarks on MILC 
configuration sets labelled in column 1 that include 
2+1 flavors of asqtad quarks (see~\cite{Davies:2009tsa} for more 
details about the ensembles). 
The results are derived from the correlators calculated in~\cite{Davies:2009tsa} 
and~\cite{newfds} along with 
Eq.(~\ref{eq:ppbardef}), but we also give results for 
additional strange quark masses on sets 1 and 2. 
$\delta x_{sea}$ is the mismatch between the sea $2m_l+m_s$ value and
the physical result divided by the physical value of $m_s$~\cite{newfds}. 
}
\label{tab:asqtadresults}
\begin{ruledtabular}
\begin{tabular}{lllllll}
Set & $\delta x_{sea}$ & $a_{\eta_s}$ (fm) & $am_{s,val}$ & $aM_{\eta_s}$ & $af_{\eta_s}$ & -$a^3 \langle \overline{\psi} \psi_s \rangle_{0}$ \\
\hline
1 & 0.47 &  0.1583(13) & 0.061 & 0.50490(36) & 0.1410(4) & 0.042399(38) \\
 & & &  0.066 & 0.52524(36) & 0.1429(4) & 0.044637(38)  \\
 & & &  0.080 & 0.57828(34) & 0.1485(4) & 0.050795(37) \\
2 & 0.91 & 0.1595(14) & 0.066 & 0.52458(35) & 0.1434(3) & 0.044714(37) \\
3 & 0.64 &  0.1247(10) & 0.0489 & 0.41133(17) & 0.1124(2) & 0.030233(14) \\
 & & &  0.0537 & 0.4310(4) & 0.1144(2) & 0.032423(15) \\
4 & 0.93 & 0.1264(11) & 0.0492 & 0.41436(23) & 0.1136(2) & 0.030585(21) \\
 & & &  0.0546 & 0.43654(24) & 0.1160(3) & 0.033041(20) \\
 & & &  0.060 & 0.45787(23) & 0.1182(4) & 0.035476(20) \\
5 & 1.5 & 0.1263(11) & 0.0491 & 0.41196(24) & 0.1135(2) & 0.030306(21) \\
 & & &  0.0525 & 0.4259(6) & 0.1149(4) & 0.031817(23) \\
 & & &  0.0556 & 0.4384(6) & 0.1161(4) & 0.033211(23) \\
6 & 0.59 & 0.0878(7) & 0.0337 & 0.29413(12) & 0.07954(9) & 0.018310(5) \\
 & & &  0.0358 & 0.30332(12) & 0.08051(9) & 0.019273(5) \\
 & & &  0.0382 & 0.31362(14) & 0.08171(15) & 0.020370(5) \\
7 & 1.1 & 0.0884(7) & 0.0336 & 0.29309(13) & 0.07959(11) & 0.018217(5) \\
 & & &  0.03635 & 0.30513(20) & 0.08095(14) & 0.019467(7) \\
8 & 0.28 & 0.0601(5) & 0.0228 & 0.20621(19) & 0.0549(2) & 0.011311(5) \\
 & & &  0.024 & 0.21196(13) & 0.0556(1) & 0.011851(3) \\
9 & 0.38 & 0.0443(4) & 0.0161 & 0.1525(2) & 0.0404(1) & 0.0075891(20) \\
\end{tabular}
\end{ruledtabular}
\end{table*}

For the first calculation we use values of the strange 
quark condensate listed in Table~\ref{tab:asqtadresults}. 
These are obtained from studies of the $\eta_s$ correlator 
on 9 different ensembles at 5 different values of the lattice spacing 
and multiple sea quark mass values. The lattice spacing 
values we use here are defined from the $\eta_s$ decay 
constant and are determined in~\cite{Davies:2009tsa}. 
From the values in Table~\ref{tab:asqtadresults} we can construct the ratio $R_s$ defined in 
Eq.~(\ref{eq:rsdef}) and fit it as a function of lattice 
spacing in exactly the same way as that described in section~\ref{subsub:fits}.
The $\mathcal{O}(\alpha_s^0)$ and $\mathcal{O}(\alpha_s^1)$ perturbative subtractions
defined in section~\ref{subsec:pt} apply here also since 
no effects appear at this order from the 
differing number of sea quarks 
or the formalism used for them 
or the improvement coefficients in the gluon action 
(the MILC 2+1 asqtad configurations 
do not include the $n_f\alpha_sa^2$ 
improvement coefficients in the gluon action). 
These effects will cause differences in the perturbation 
theory at 
$\mathcal{O}(\alpha_s^2)$. 
The $\alpha_s^2$ and higher order divergent pieces of the 
condensate are included 
in the fit with coefficients that, as before, are 
given a prior value of 0(4). The appropriate $\alpha_s$ 
value in this case is  
$\alpha_{V}^{n_f=3}(2/a)$ 
rather than $\alpha_V^{n_f=4}(2/a)$. 
Multiple valence $s$ quark masses are given at each lattice spacing 
and we allow for linear and quadratic dependence on 
the mistuning of the $s$ quark mass, again as described 
in section~\ref{subsub:fits}. We allow for mistuning of the 
sea quark masses through use of the parameter $\delta x_{sea}$~\cite{newfds}. 

\begin{center}
\begin{figure}[ht]
\includegraphics[width=0.9\hsize]{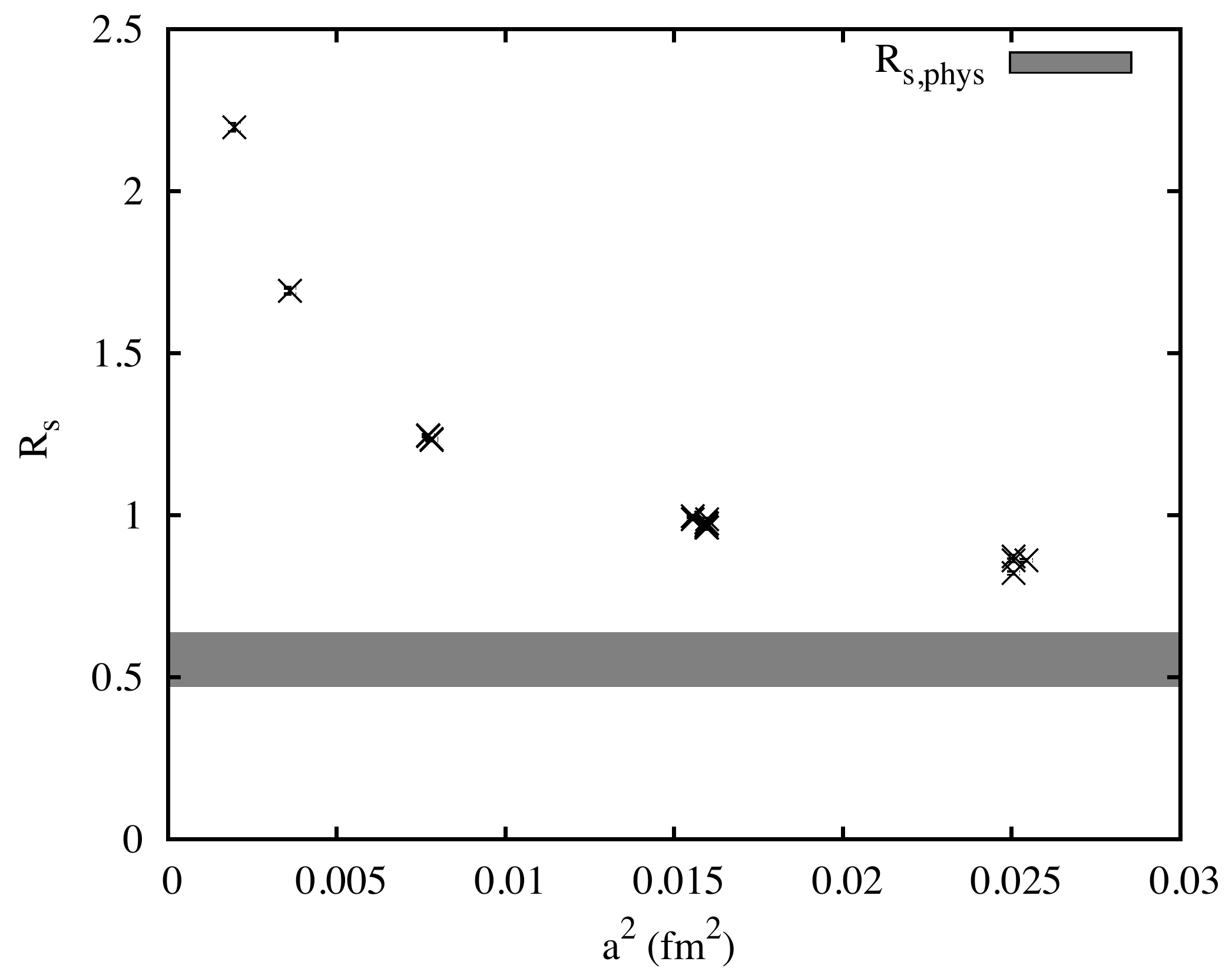}
\caption{Results from fitting the values for 
$R_s$ obtained on MILC configurations including
2+1 flavors of asqtad sea quarks, as described 
in the text. The crosses give the calculated 
values after perturbative subtraction 
through $\mathcal{O}(\alpha_s)$. 
The hatched band corresponds to the 
fitted physical value after removing the 
remaining power divergence and discretisation and 
sea quark mass effects. Compare to Fig.~\ref{fig:rhisqfit} 
which includes 2+1+1 flavors of HISQ sea quarks. }
\label{fig:asqtadfit}
\end{figure}
\end{center}

Figure~\ref{fig:asqtadfit} shows, as a hatched band, the physical 
result from the fit, which has $\chi^2/\mathrm{dof} = 0.4$ 
for 20 degrees of freedom. 
For comparison the data points 
given are the values after perturbative subtraction 
through $\mathcal{O}(\alpha_s)$. 
The physical value obtained is 
\begin{equation}
R_{s,phys} = 0.555(84).
\label{eq:asqtadval}
\end{equation}
This is completely consistent with the result from 2+1+1 flavors of 
HISQ sea quarks in section~\ref{subsub:fits}, and has a similar error.  
It is not such a complete calculation, lacking light quark mass results 
and not having such light sea quark masses, and is therefore not our 
preferred final result. It provides a strong check of our 2+1+1 result, 
however, being a completely independent set of numbers. 
The fits to the 2+1 results give very similar behaviour to that seen for the 
2+1+1 case, for example choosing a coefficient of 
the $\alpha_s^2/a^2$ divergence of around 2. 

For the second calculation we use values of the strange condensate 
from the HOTQCD collaboration~\cite{Bazavov:2011nk}. 
They generated ensembles with an improved gluon 
action and $u/d$ and $s$ quarks in the sea using the HISQ 
formalism. The QCD action differs slightly from 
that in section~\ref{sec:lattice}. 
Apart from missing $c$ quarks in the sea, the gauge 
field configurations here 
are improved through $\mathcal{O}(a^2)$ 
at tree-level and without tadpole-improvement, 
i.e. the fairly substantial $\mathcal{O}(\alpha_sa^2)$ 
improvement coefficients were not included. The 
lattice spacing was determined using the 
$r_1$ heavy quark potential parameter, or $r_0$ 
on the coarsest lattices where $r_1/a < 2$.  
The $s$ quark mass was tuned by determining the mass 
of the $\eta_s$ meson and the light quark mass 
was taken as $m_s/20$ (with some values available 
for $m_s/5$ but we have not used those). 
Results for the zero temperature strange condensate 
are available at 24 values of the lattice spacing from 
0.2 fm to 0.07fm (Table 14 of~\cite{Bazavov:2011nk} gives values 
for two times the condensate). Note that these results 
were obtained by direct calculation of 
$\langle \mathrm{Tr}M^{-1} \rangle$ 
using stochastic techniques.  
Corresponding values of the lattice spacing are given 
in Table 16 of~\cite{Bazavov:2011nk}; some missing values 
can be inferred from the tables of temperature values 
at the corresponding value of $\beta$. 

\begin{center}
\begin{figure}[ht]
\includegraphics[width=0.9\hsize]{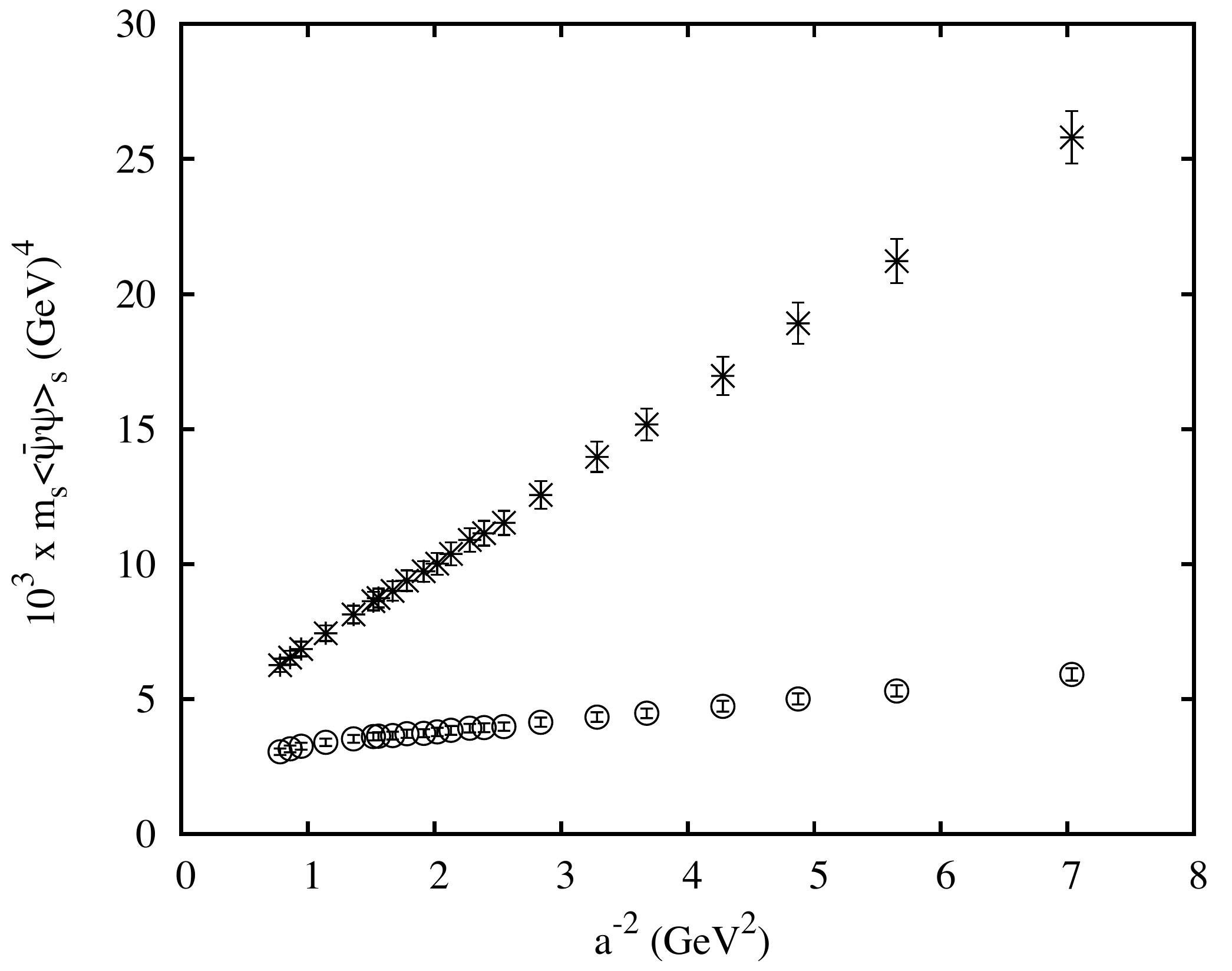}
\caption{Bursts give the raw data for 
$m_s\langle \overline{\psi} \psi_s \rangle$ 
from~\cite{Bazavov:2011nk} as a function of the square 
of the inverse lattice spacing. 
Open circles give results after subtraction 
of the $\mathcal{O}(\alpha_s)$ perturbative 
contribution.} 
\label{fig:hotqcdraw}
\end{figure}
\end{center}

Figure~\ref{fig:hotqcdraw} shows the raw unsubtracted results for 
$m_s \langle \overline{\psi} \psi_s \rangle$  as a function 
of the square of the inverse lattice spacing, 
as well as the values after making the complete 
subtraction through $\mathcal{O}(\alpha_s)$ as given in 
Eq.~(\ref{eq:deltapert}).  
As for $R_s$ in section~\ref{subsub:firstlook} 
(Fig.~\ref{fig:unsub}), the unsubtracted results show clear evidence 
of a quadratic term in $a^{-1}$ which is significantly 
reduced, but not completely absent, after the perturbative 
subtraction. 

For a subset of 9 lattice spacing values meson masses and 
decay constants are also given in Tables 18 and 
19 of~\cite{Bazavov:2011nk}. In fact we use the 
7 finest values only because the $s$ quark mass 
is not as well-tuned for our purposes 
on the coarsest two lattices. Note that the decay 
constant values need to be multiplied 
by $\sqrt{2}$ to match the convention used here. 
For these we can 
construct the ratio $R_s$ given in~\ref{eq:rsdef}.   
To obtain a physical result for 
$R_s$ 
we fit the subtracted results as a function of lattice 
spacing in the same way
as in section~\ref{subsub:fits}, apart from  
the use of $\alpha_{V}^{n_f=3}(2/a)$ 
rather than $\alpha_V^{n_f=4}(2/a)$. 

The physical value for $R_s$ obtained from the fit is 
\begin{equation}
R_s = 0.79(34)
\label{eq:hotqcdval}
\end{equation}
This is much less accurate than the result from section~\ref{sec:lattice}, 
but agrees both with that and the result from the 
MILC 2+1 asqtad ensembles given earlier in this section. 
We have not extracted a light quark condensate from 
the HOTQCD results because finite volume sensitivity 
obscures the power divergence and leads to larger errors. 

We conclude from this that there is no sign of disagreeement 
between the strange quark condensate extracted with $u$, $d$ and 
$s$ quarks in the sea and those including also $c$ quarks in the sea. 

\bibliography{cond}

\end{document}